\documentclass[aps,10pt,pra,superscriptaddress,showpacs,floatfix,notitlepage,longbibliography,twocolumn]{revtex4-1}
\usepackage{graphicx}
\usepackage{amsmath}
\usepackage{amsthm}
\usepackage{amssymb}
\usepackage[usenames]{color}
\usepackage{hyperref}
\RequirePackage[normalem]{ulem}
\hypersetup{
    colorlinks=true,       
    linkcolor=cyan,          
    citecolor=magenta,        
    filecolor=magenta,      
    urlcolor=cyan,           
    runcolor=cyan
}

\usepackage{cancel}
\usepackage{bm}
\usepackage{threeparttable}
\usepackage{subfigure}
\usepackage{upgreek }
\usepackage{marginnote}

\newcommand{\beq}{\begin{equation}}
\newcommand{\eeq}{\end{equation}}
\newcommand{\beqnn}{\begin{equation*}}
\newcommand{\eeqnn}{\end{equation*}}
\newcommand{\bea}{\begin{eqnarray}}
\newcommand{\eea}{\end{eqnarray}}
\newcommand{\beann}{\begin{eqnarray*}}
\newcommand{\eeann}{\end{eqnarray*}}
\newcommand{\bes} {\begin{subequations}}
\newcommand{\ees} {\end{subequations}}

\newcommand{\ket}[1]{ | #1\rangle}
\newcommand{\bra}[1]{\langle #1 | }

\newcommand{\ident}{\openone}

\newcommand{\1}{\openone}

\newcommand{\ignore}[1]{}

\begin{document}
\title{Validating a Two Qubit Non-Stoquastic Hamiltonian in Quantum Annealing}
\author{Tameem Albash}
\affiliation{Department of Electrical and Computer Engineering, Department of Physics and Astronomy, and Center for Quantum Information and Control, CQuIC, University of New Mexico, Albuquerque, New Mexico 87131, USA}

\begin{abstract}
\noindent We propose a two qubit experiment for validating tunable antiferromagnetic $XX$ interactions in quantum annealing.  Such interactions allow the time-dependent Hamiltonian to be non-stoquastic, and the instantaneous ground state can have negative amplitudes in the computational basis.  Our construction relies on how the degeneracy of the Ising Hamiltonian's ground states is broken away from the end point of the anneal: above a certain value of the antiferromagnetic $XX$ interaction strength, the perturbative ground state at the end of the anneal changes from a symmetric to an antisymmetric state.  This change is associated with a suppression of one of the Ising ground states, which can then be detected using solely computational basis measurements. We show that a semiclassical approximation of the annealing protocol fails to reproduce this feature, making it a candidate `quantum signature' of the evolution.
\end{abstract}

\maketitle

\section{Introduction}
%
There remain no demonstrated examples of a quantum speedup using quantum annealing \cite{finnila_quantum_1994,Brooke1999,kadowaki_quantum_1998,Farhi:01,Santoro} outside of the oracular setting \cite{Roland:2002ul,Somma:2012kx}.  Evidence that is often cited for this predicament is that standard quantum annealing implements a stoquastic Hamiltonian \cite{Bravyi:2009sp,Bravyi:QIC08,Bravyi:2014bf,Marvian2019,Klassen2019twolocalqubit,Klassen2019b} throughout the anneal, 
and ground state adiabatic quantum computing \cite{Farhi:00} with a stoquastic Hamiltonian is not expected to be universal \cite{Bravyi:QIC08}.  The restriction to stoquastic Hamiltonians, for which a classical probability distribution can be associated with the ground state, also makes the annealing protocol amenable to classical simulation using Quantum Monte Carlo (QMC) techniques.  While QMC does not simulate the unitary dynamics of quantum annealing, it does reproduce the scaling dependence with the minimum gap for certain problems \cite{2015arXiv151008057I}, but crucially it fails to do so for others \cite{Hastings:2013kk,Andriyash:2017aa}, making it unclear a priori for which classes of problems quantum annealing can provide a legitimate scaling advantage over QMC.

The introduction of novel interactions at intermediate points in the anneal that make the Hamiltonian non-stoquastic would hinder QMC techniques from being efficient simulators of the annealing protocol because of the associated sign-problem \cite{PhysRevLett.94.170201,Marvian2019,Klassen2019twolocalqubit,Klassen2019b}. This effectively eliminates QMC as a legitimate classical competitor, and it remains an open question to what extent this will improve the current situation of demonstrating a quantum speedup using quantum annealing \cite{PhysRevA.95.042321,Nishimori:2016aa,crosson2014different,Hormozi:2016aa,PhysRevA.99.042334} \footnote{We note that tunable non-stoquastic interactions are necessary for universal adiabatic quantum computing \cite{Biamonte:07}, although we are not considering this question here.}.

Nevertheless, experimental realizations of such interactions are ongoing \cite{DWaveNonStoq,Kerman_2019}, and here we propose a method to validate the implementation of tunable antiferromagnetic $XX$ interactions within the constraints of the quantum annealing protocol, i.e. the evolution terminates on a Hamiltonian $H_{\mathrm{P}}$ that is diagonal in the computational basis, and only measurements in the computational basis at the end of the anneal are allowed. 

In order to accomplish this objective, our construction relies on a change in the ground state of the time-dependent Hamiltonian near the end of the anneal. As the strength of the antiferromagnetic $XX$ interaction is increased, the ground state changes from a symmetric combination of the three ground states of $H_\mathrm{P}$ to an antisymmetric combination of two of the three ground states of $H_{\mathrm{P}}$.  The suppression of population in one of the ground states is then a measurable signature of this transition.  The proposal thus tests both the tunability of the interactions and the ability of the quantum annealer to implement a ground state with non-trivial relative phases between the computational basis states.  We demonstrate that at least one semiclassical model of quantum annealing, whereby qubits are replaced by spin vectors, fails to reproduce this signature.

Our paper is structured as follows.  In Sec.~\ref{Sec:Perturbation}, we show how perturbation theory predicts the breaking of the degeneracy of the final three ground states to give a unique ground state near the end of the anneal.  In Sec.~\ref{sec:Dynamics}, we present the results of dynamical simulations to test the robustness of the proposal to several noise models.  In Sec.~\ref{sec:Semiclassical}, we provide a comparison to a semiclassical model of quantum annealing, and we conclude in Sec.~\ref{sec:Conclusion}.
\section{Perturbation Theory} \label{Sec:Perturbation}
We begin by considering the following time-dependent 2-qubit Hamiltonian:
\begin{eqnarray} \label{eqt:IdealH}
H(s)/(\hbar \omega) &=& -(1-s) \left( \sigma_1^x + \sigma_2^x  \right) + \alpha s (1-s) \sigma_1^x \sigma_2^x  \nonumber \\
&& + s \left(- \sigma_1^z - \sigma_2^z + \sigma_1^z \sigma_2^z \right) \ ,
\end{eqnarray}
where $\sigma_i^\gamma$ is the Pauli-$\gamma$ operator acting on qubit $i$, $|\alpha|$ is the strength of the $XX$ interaction with $\alpha < 0$ or $> 0$ corresponding to a ferromagnetic or antiferromagnetic interaction, and $\hbar \omega$ sets the overall energy scale.  Here $H_\mathrm{P}$ is given by the last term in parenthesis.  The Hamiltonian is invariant under the interchange of the two qubits, so all energy eigenstates will be invariant up to an overall phase under the nterchange of the two qubits. The ground state of the Ising Hamiltonian at $s=1$ exhibits a three-fold degeneracy: $\ket{00}, \ket{01}, \ket{10}$, where $\ket{0}$ is the eigenstate of $\sigma^z$ with positive eigenvalue.  

If we denote the perturbation parameter by $\Gamma = 1-s$, expanding $H(s)$ around $s = 1$ gives to first order $H(s) = H(1) + \hbar \omega \Gamma  V_1$, where
\beq
V_1 =  - (\sigma_1^x + \sigma_2^x) + \alpha \sigma_1^x \sigma_2^x - \left(- \sigma_1^z - \sigma_2^z + \sigma_1^z \sigma_2^z \right) \ ,
\eeq
When $V_1$ is projected onto the ground state subspace of $s=1$, the three eigenstates with their corresponding eigenvalues of the resulting operator are given by:
\bes
\begin{align}
\ket{\lambda}  &= \frac{1}{\sqrt{2}} \left( \ket{01} - \ket{10} \right)  , \  \lambda = 1 - \alpha \ , \\
\ket{\lambda'} & = \frac{1}{\sqrt{ \gamma_+^2 + 2}} \left(\gamma_+ \ket{00} + \ket{01} + \ket{10} \right),  \ \lambda' = 1 + \gamma_-  ,\\
\ket{\lambda''} & = \frac{1}{\sqrt{ \gamma_-^2 + 2}} \left(\gamma_- \ket{00} + \ket{01} + \ket{10} \right), \ \lambda'' = 1 + \gamma_+  , 
\end{align}
\ees
with $\gamma_\pm = \frac{1}{2} \left( \alpha \pm \sqrt{8 + \alpha^2} \right)$.  The states $\ket{\lambda'}$ and $\ket{\lambda''}$ are symmetric under the interchange of the qubits, whereas $\ket{\lambda}$ is antisymmetric under the interchange of the qubits and does not have any weight on the $\ket{00}$ state.  A classical thermal state at $s=1$ will by definition have equal populations on all three ground states, approaching the value $1/3$ in the limit of zero temperature.

The state with smallest eigenvalue $\left\{\lambda, \lambda' \right\}$ is the ground state for $\Gamma = 0^+$, so for $\alpha < 1$ the ground state is $\ket{\lambda'}$ and for $\alpha > 1$, the ground state is $\ket{\lambda}$.  As $\alpha$ is swept through this point, the ground state changes from a symmetric to an antisymmetric state.  Quantitatively, we can consider the expectation value of the SWAP operator, $\mathrm{SWAP} = \frac{1}{2} \left( \ident + \sum_{\gamma = \left\{x,y,z \right\}} \sigma_1^\gamma \sigma_2^\gamma \right)$, which changes from $+1$ to $-1$ as we cross $\alpha = 1$. 

Because the ground state at $s=0$ is symmetric, for $\alpha > 1$ there is a true level-crossing in the spectrum associated with this change in ground state.
This level crossing is predicted already by considering the perturbative corrections to the energies associated with the states $\ket{\lambda}$ and $\ket{\lambda'}$ at second-order in perturbation theory.  The non-zero contributions are given by:
\bes
\begin{align}
\frac{E_{(2)}(\Gamma)}{\hbar \omega} =& -1 + \bra{\lambda} V_1 \ket{\lambda} \Gamma + \bra{\lambda}V_2 \ket{\lambda} \Gamma^2 \ , \\
\frac{E_{(2)}'(\Gamma)}{\hbar \omega}  =& -1 + \bra{\lambda'} V_1 \ket{\lambda'} \Gamma \nonumber \\
& + \left( \bra{\lambda'} V_2 \ket{\lambda'} - \frac{1}{4} \left| \bra{\lambda'} V_1 \ket{11} \right|^2 \right)\Gamma^2 \ ,
\end{align}
\ees
where $V_2 = -\alpha \sigma_1^x \sigma_2^x$ is a second-order perturbation that also contributes. For $\alpha > 0$, $E_{(2)}$ curves upwards and $E_{(2)}'$ curves downwards, but only for $\alpha > 1$ does  $E_{(2)}$ reach a lower value than $E_{(2)}'$ for $s < 1$, which leads to the two energies crossing at some intermediate $s$ value.
%
\section{Dynamics} \label{sec:Dynamics}
\begin{figure}[htbp] 
   \centering
   \subfigure[]{\includegraphics[width=0.48\columnwidth]   {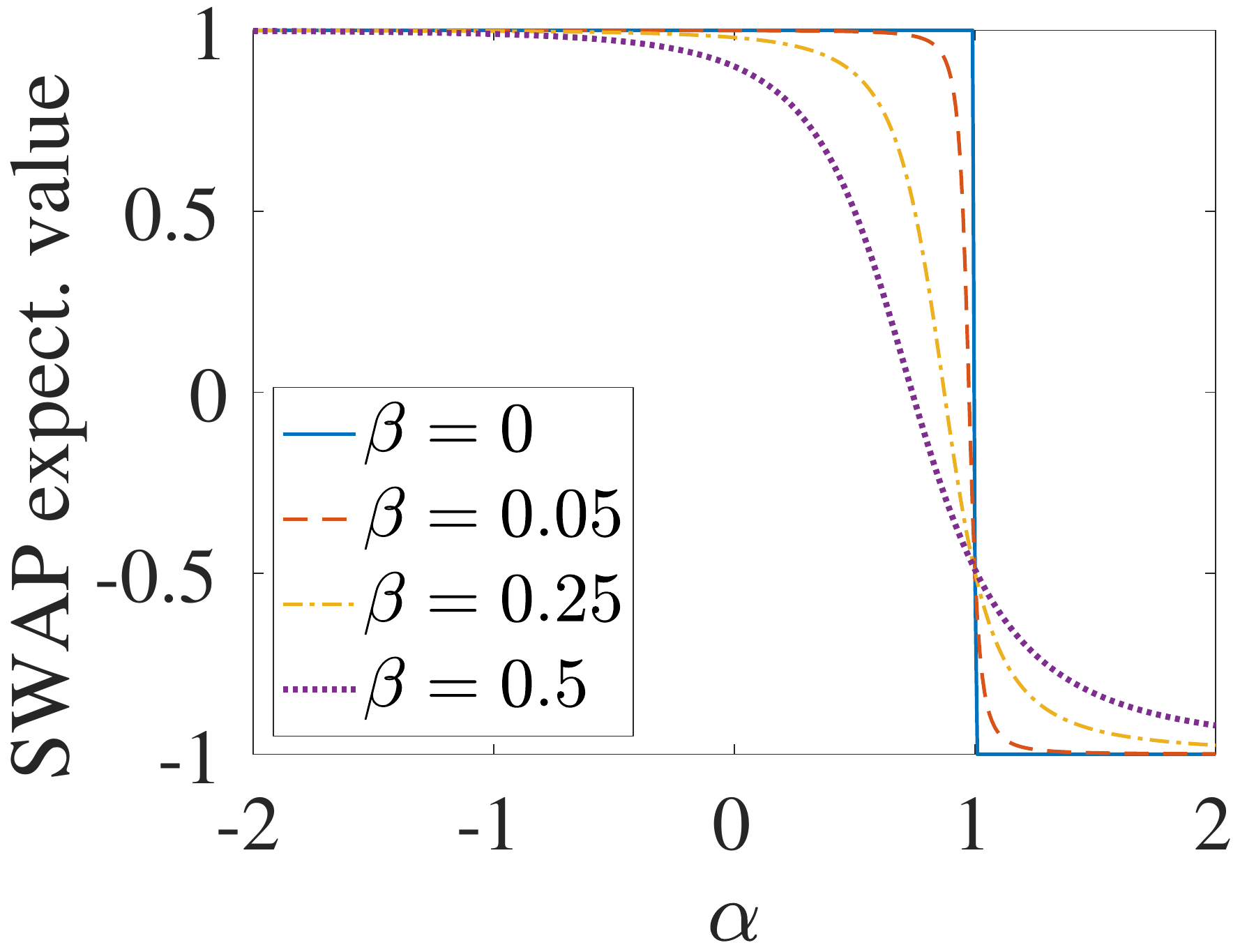} \label{fig:2qGadgetA}}
   \subfigure[]{\includegraphics[width=0.48\columnwidth]{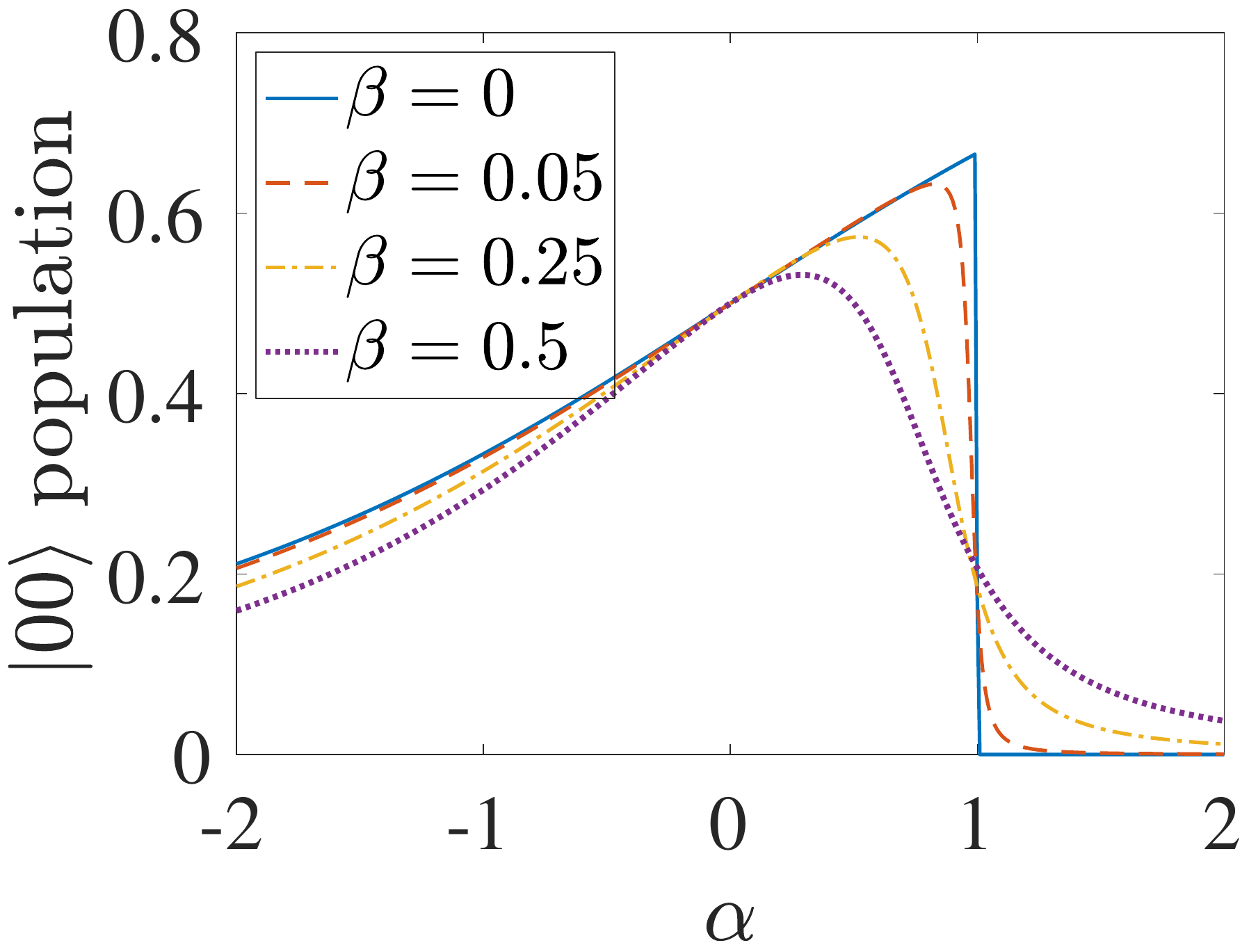} \label{fig:2qGadgetB}}
   \caption{Properties of the perturbative ground state calculated from first order perturbation theory for various $\beta$ values.  (a) Expectation value of the SWAP operator. (b) Population in the state $\ket{00}$.} \label{fig:2qGadget}
\end{figure}

%
Because the energy level crossing discussed above is associated with a symmetry of the Hamiltonian, an adiabatic evolution will not follow the global ground state through the crossing.  Therefore, we propose to break the qubit permutation symmetry by offsetting one of the transverse fields:
\begin{eqnarray} \label{eqt:PerturbedH}
\frac{H(s)}{\hbar \omega} &=& -(1-s) \left( \sigma_1^x + (1 - \beta) \sigma_2^x  \right) + \alpha s (1-s) \sigma_1^x \sigma_2^x \nonumber \\
&& + s \left(- \sigma_1^z - \sigma_2^z + \sigma_1^z \sigma_2^z \right) \ ,
\end{eqnarray}
with $\beta > 0$.  The true level-crossing associated with $\alpha > 1$ now becomes an avoided level crossing, and an adiabatic evolution is able to follow the global ground state throughout the anneal.  The change in the character of the ground state remains apparent for a range of $\beta$ values as $\alpha$ is tuned, as we show in Fig.~\ref{fig:2qGadget}.  The transition at $\alpha = 1$ becomes sharper as $\beta \to 0$.

We now consider the robustness of our proposal to several noise models.  We first consider the effect of interactions with a thermal environment as described by a weak-coupling limit master equation \cite{ABLZ:12-SI}.  We give details of this master equation in Appendix \ref{App:ME}. We show in Fig.~\ref{fig:OpenSystem1} the behavior as a function of the dimensionless system-bath coupling $\kappa^2$.  We see that for $\alpha = 2$, the closed system evolution ($\kappa^2 = 0$) is close to adiabatic for $\omega t_f \gtrsim 5 \times 10^3$.  As soon as we have a non-zero system bath interaction, the population of the $\ket{00}$ state approaches the classical Ising Gibbs state population of $1/3$ in the long time limit, as expected by the open-system adiabatic theorem \cite{PhysRevA.93.032118}.  If $\kappa^2$ is sufficiently small, we observe the expected competition between closed and open system adiabaticity: closed-system adiabaticity sends the population in the $\ket{00}$ state to zero, whereas open-system adiabaticity sends the population to $1/3$.  For sufficiently small $\kappa^2$ the closed system adiabatic time scale is smaller than the open system time scale, and we can continue to observe our desired signature of population suppression of the $\ket{00}$ state before open-system effects begin to dominate. 

We also consider the effect of implementation errors in the parameters of $H_\mathrm{P}$ to our population suppression signature. We model these as independent Gaussian random variables with zero mean and standard deviation $\sigma$. In Fig.~\ref{fig:OpenSystem2}, we show the populations in the computational basis states as a function of $\sigma$, where we again observe that there is a noise threshold before the distribution looks classical.  We emphasize that the evolution for low $\sigma$ is non-adiabatic, even though we picked a time-scale that is close to adiabatic for the original Hamiltonian. This is critical since a random noise instantiation picks one of the three ground states to be the new ground state with equal probability, so in the adiabatic limit of the noisy Hamiltonian, we recover the classical result upon averaging over noise realizations.  Therefore, we find that we can continue to observe our desired signature of population suppression if there is a separation between the noise-less and noisy adiabatic time scales.

\begin{figure}[htbp]
\begin{center}
\subfigure[]{\includegraphics[width=0.48\columnwidth]{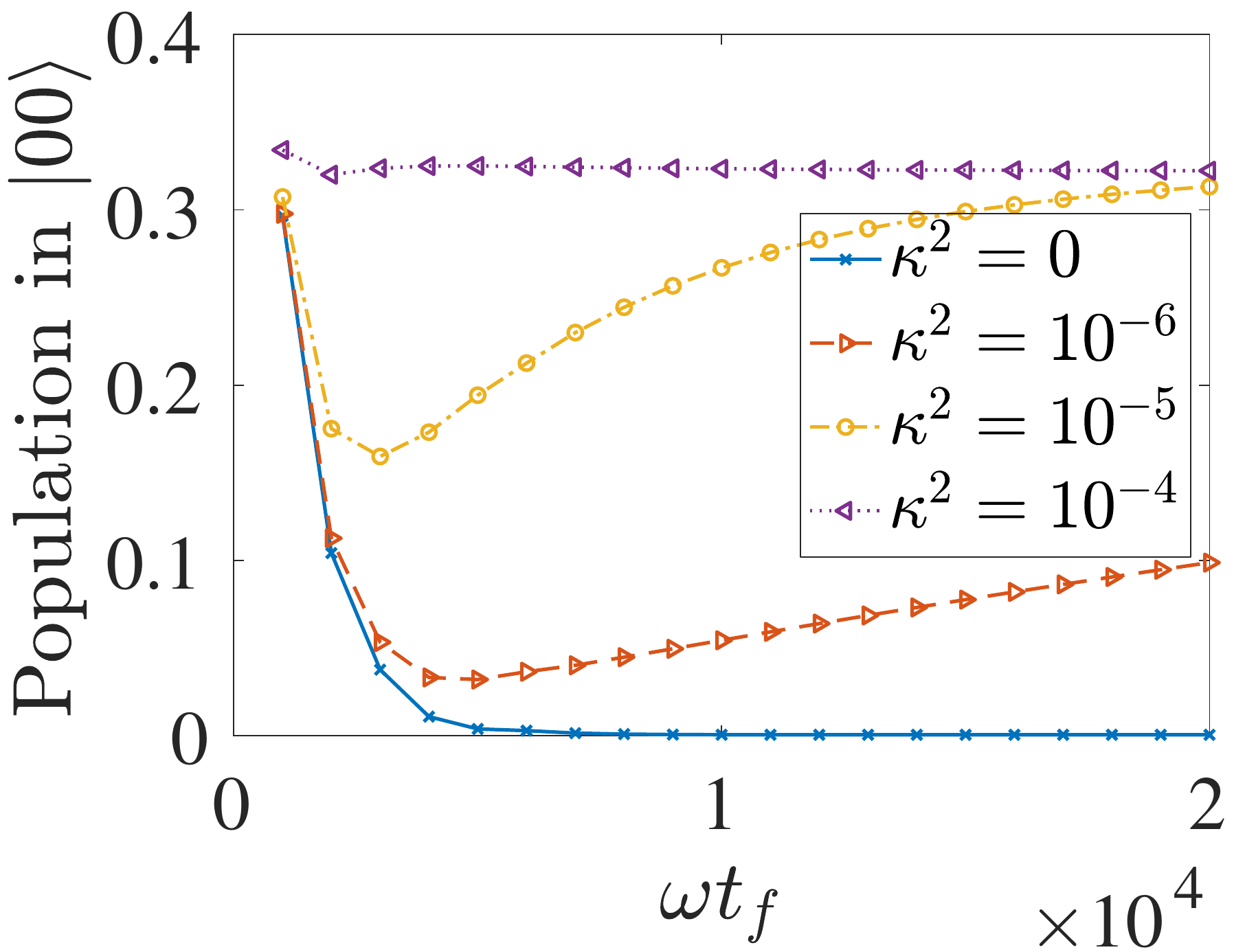} \label{fig:OpenSystem1}}
   \subfigure[]{\includegraphics[width=0.48\columnwidth]{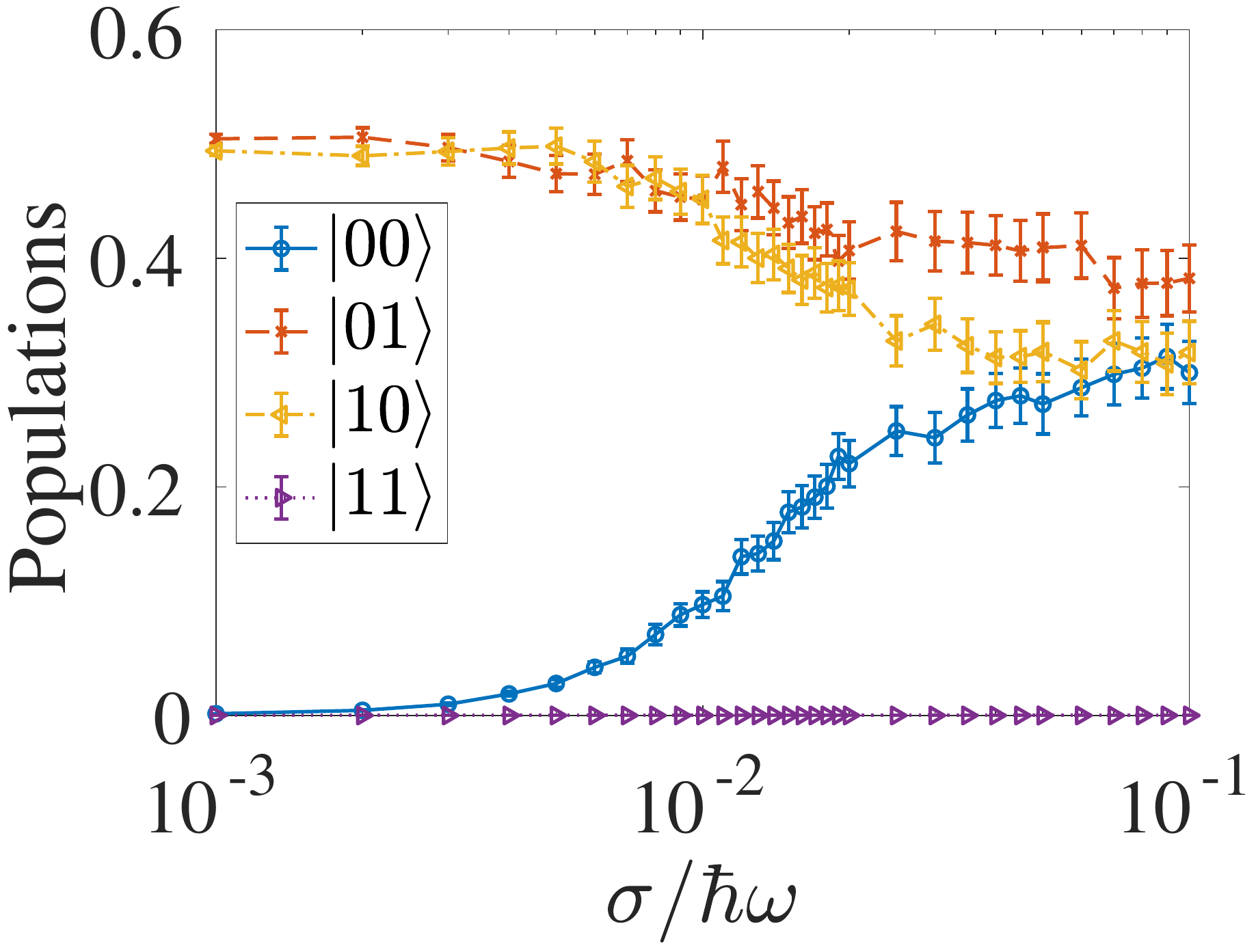}  \label{fig:OpenSystem2}}
   \caption{(a) Population in the $\ket{00}$ state at the end of the anneal for a system evolving according to the weak-coupling master equation simulations with $\alpha = 2, \beta = 0.05$ , dimensionless system-bath coupling $\kappa^2$, energy scale $k_B T / \hbar \omega = 1.57$, and varying total annealing time $t_f$. (b) Computational basis populations of the evolved state at the end of the anneal for $\beta = 0.05$, $\omega t_f = 10^4$, $\kappa^2 = 0$,$\alpha = 2$ with Gaussian noise $\mathcal{N} \sim (0, \sigma^2)$ on the Ising parameters. Results are for the mean value of $10^3$ independent noise realization, and the error bars correspond to the $95\%$ confidence interval calculated using a bootstrap. }
\end{center}
\end{figure}
%
\section{Semiclassical Analogue} \label{sec:Semiclassical}
\begin{figure}[htbp]
\begin{center}
\subfigure[]{\includegraphics[width=0.48\columnwidth]{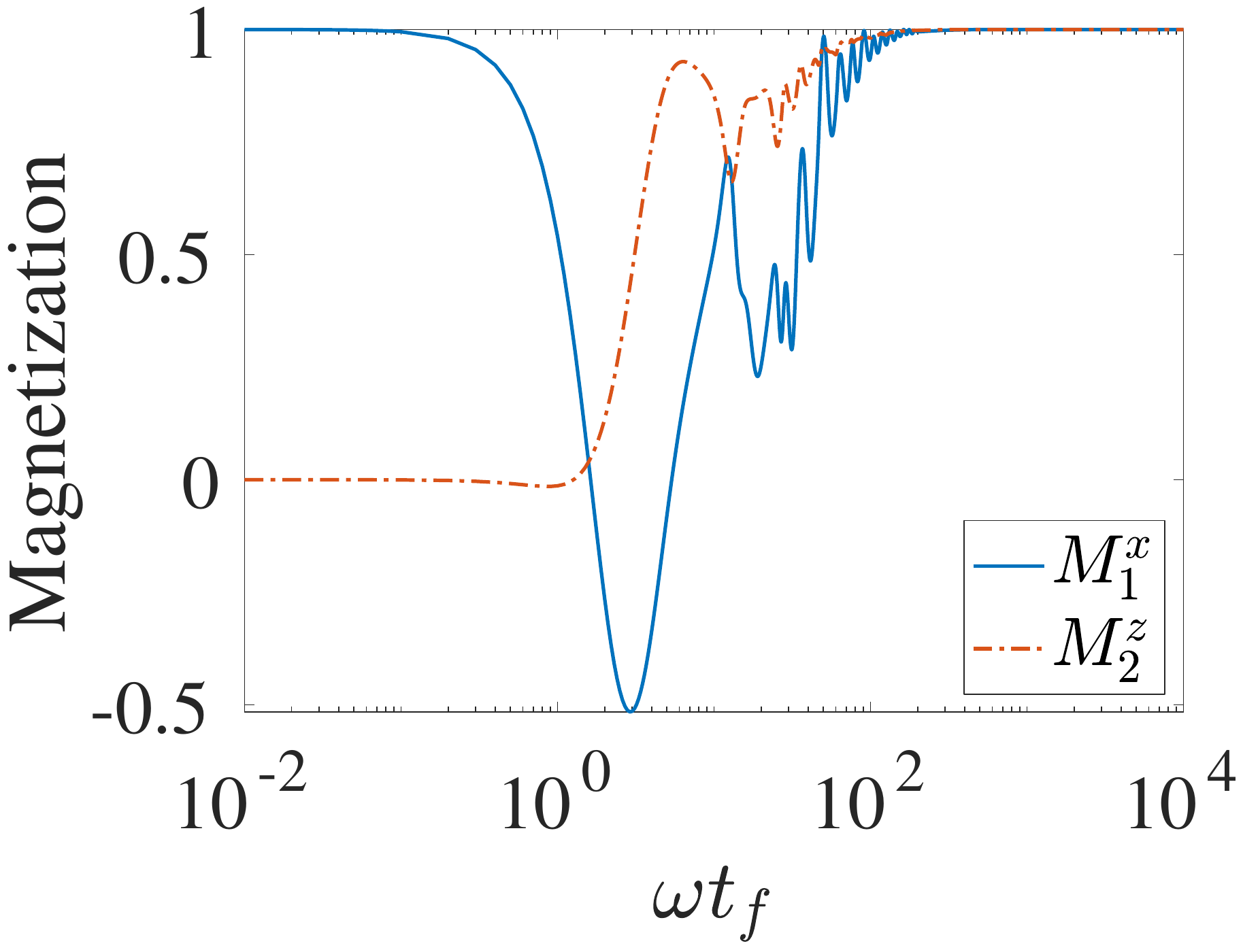} \label{fig:SVD1}}
\subfigure[]{\includegraphics[width=0.48\columnwidth]{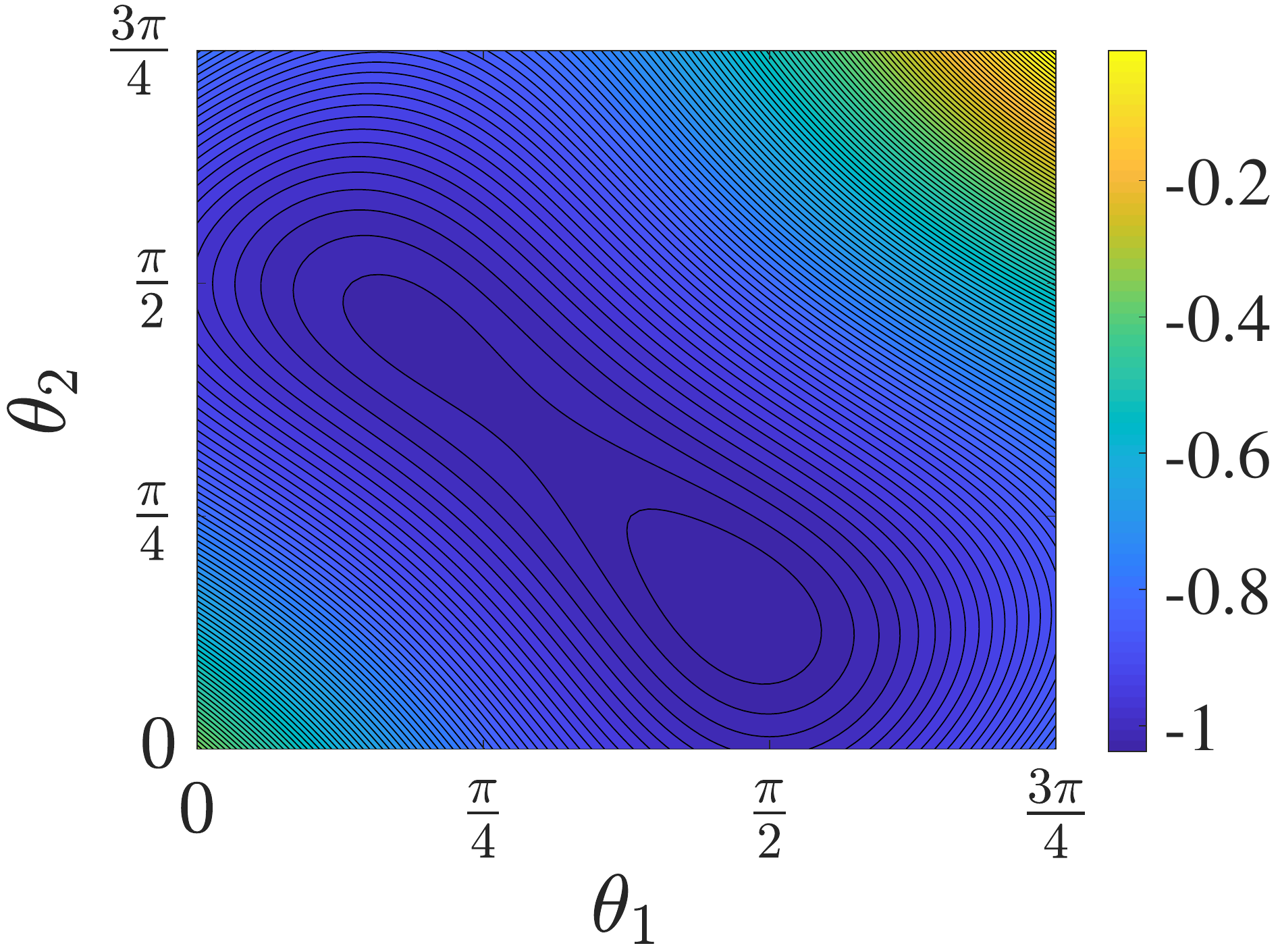} \label{fig:SVD2}}
   \caption{(a) Spin-vector dynamics results for $\alpha = 2$ and $\beta = 0.05$. (b) Cut through the semiclassical potential for fixed $\phi_i = 0$ for the system parameters  $\alpha = 2, \beta = 0.05, s=0.35$.}  \label{fig:SVD}
\end{center}
\end{figure}

We consider a semiclassical limit of qubits corresponding to the saddle-point approximation of the path integral for the qubit system \cite{klauder1979path}, where the qubits are replaced by classical spin vectors $\vec{M}_i = \left( \sin \theta_i \cos \phi_i, \sin \theta_i \sin \phi_i, \cos \theta_i \right)$ corresponding to the average magnetization of a spin-coherent state \cite{Radcliffe_1971} along the $(x,y,z)$ directions.
The dynamics of the spin vectors are given by:
\beq
\frac{d}{ds} \vec{M}_i = \omega t_f  \vec{H}_i \times \vec{M}_i \ ,
\eeq
where
\bes
\begin{align}
\vec{H}_1 & = 2 \left( -(1-s) +  \alpha s(1-s) M_2^x \right) \hat{x} + 2 s \left( - 1 + M_2^z \right) \hat{z} \ , \\
\vec{H}_2 & = 2 \left( -(1-s)(1-\beta) +  \alpha s(1-s) M_1^x \right) \hat{x} \nonumber  \\
& \  + 2 s \left( - 1 + M_1^z \right) \hat{z} \ .
\end{align}
\ees
(For a detailed derivation, see Ref.~\cite{owerre2015macroscopic,q-sig2}.) 
We show in Fig.~\ref{fig:SVD1} the resulting magnetization at $s=1$ for increasing $\omega t_f$, where we see the system approaches the values $\vec{\theta} = (0,\pi/2)$ at very long times. 

In order to understand the long time behavior, it is useful to consider the potential energy landscape on which the dynamics occurs:
\begin{eqnarray} \label{eqt:SVMCPotential}
V(s)&=& -(1-s) \left( M_1^x+ (1 - \beta) M_2^x  \right) + \alpha s (1-s)M_1^x M_2^x \nonumber \\
&& + s \left(-M_1^z - M_2^z + M_1^z M_2^z \right) \ ,
\end{eqnarray}
which amounts to replacing the Pauli operators in Eq.~\eqref{eqt:PerturbedH} by $\sigma_i^x \mapsto M_i^x , \sigma_i^z \mapsto M_i^z$.  For $\beta = 0$ and $\phi_i = 0$, the two dimensional potential energy landscape as a function of $\vec{\theta}$ exhibits the features we associate with second-order phase transitions as we increase $s$. Specifically, a single global minimum bifurcates into two equal in energy minima, which move towards the positions $\vec{\theta} = (\pi/2, 0)$ and $(0, \pi/2)$ respectively as $s$ goes to 1, irrespective of the value of $\alpha$. For finite $\beta$, one of the two minima is always energetically favored (shown in Fig.~\ref{fig:SVD2}), and for a sufficiently slow evolution, the system follows this minimum. 

\begin{figure}[htbp]
\begin{center}
\subfigure[]{\includegraphics[width=0.48\columnwidth]{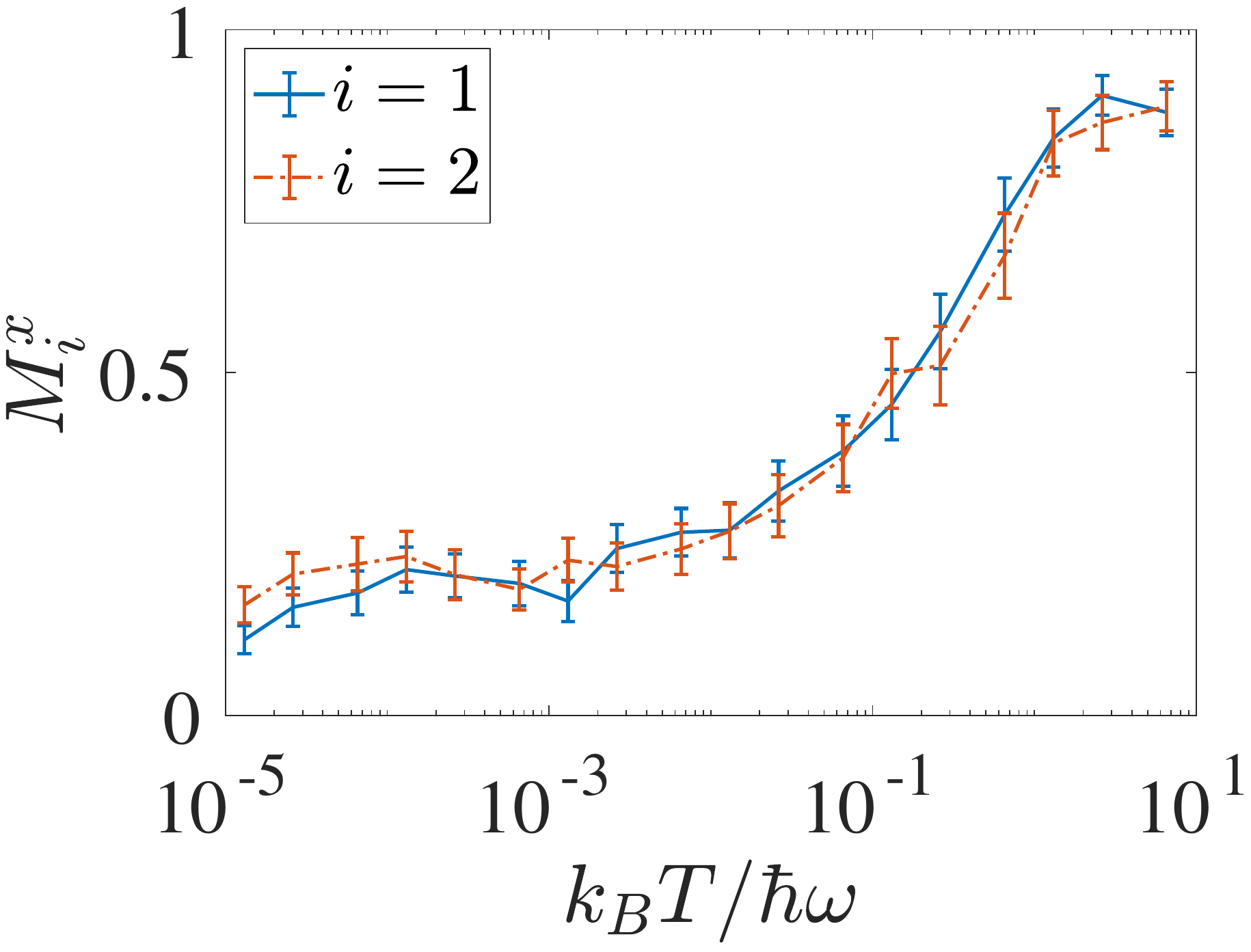} \label{fig:SVMC_Mx}}
\subfigure[]{\includegraphics[width=0.48\columnwidth]{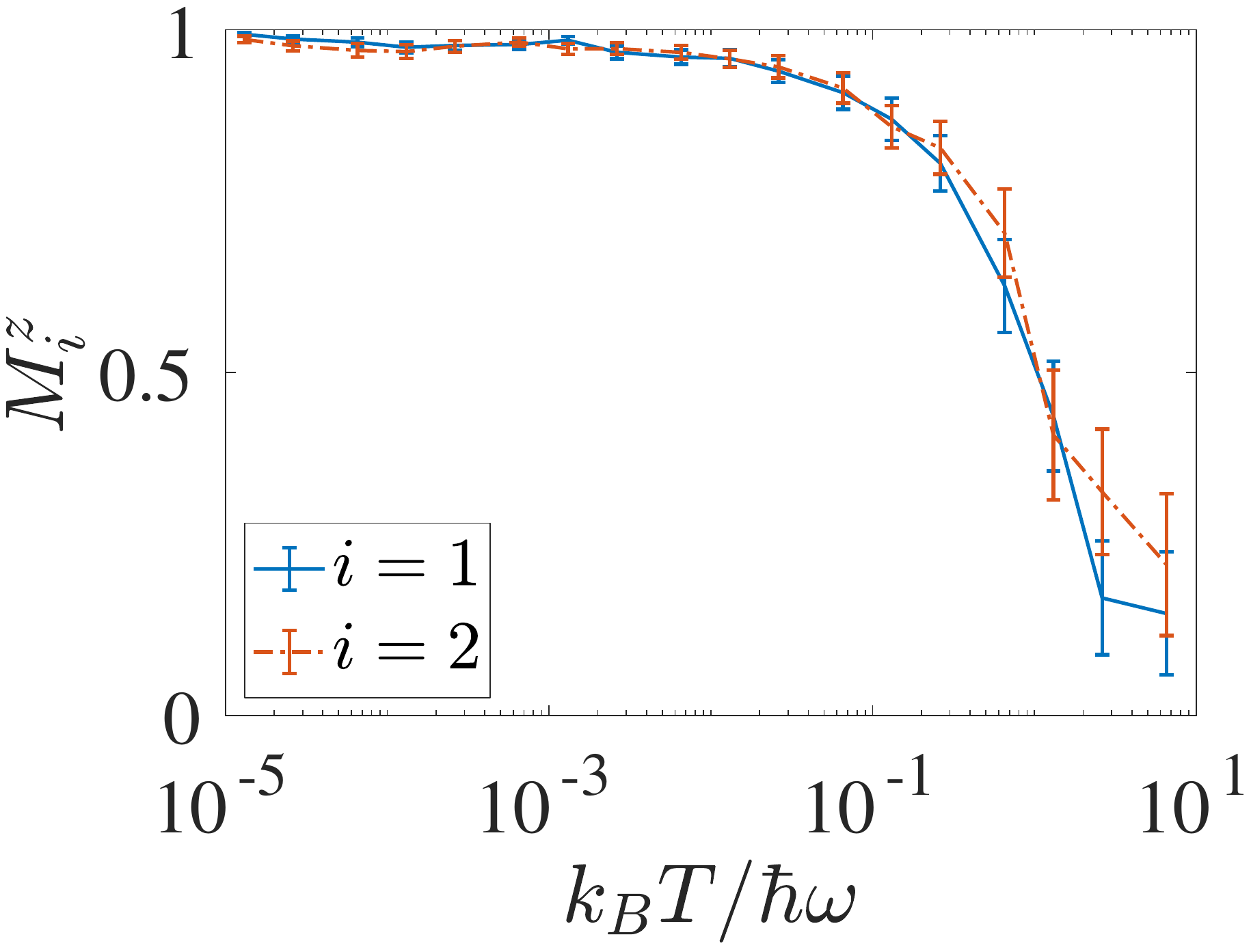} \label{fig:SVMC_Mz}}
   \caption{Spin-vector Monte Carlo results for varying temperatures at $\alpha = 2, \beta = 0.05$. (a) Average magnetization along $x$-axis.  (b) Average magnetization along $z$-axis. The  simulations use $10^6$ sweeps for each of the $10^4$ independent runs performed.  The error bars correspond to $2\sigma$ confidence intervals generated by $10^3$ bootstraps over the independent runs. } \label{fig:SVMC}
\end{center}
\end{figure}

We can also consider an evolution of the spin vectors described by Monte Carlo updates, whereby a random $\vec{M}_i$ is drawn, and the update is accepted according to the Metropolis-Hastings criteria \cite{Metropolis1953,HASTINGS01041970} using the time-dependent potential in Eq.~\eqref{eqt:SVMCPotential}.  Further details are given in Appendix \ref{App:SVMC}.  This evolution can be thought of as including the effect of a finite temperature environment in the strong coupling limit \cite{PhysRevA.94.062106}, although here we do not restrict the evolution to a plane.  We show in Fig.~\ref{fig:SVMC} the dependence on temperature, where we see a strong bias for the $M_1^z = M_2^z =1$ state at low temperatures.  To understand the reason for this bias, we expand the potential at $s=1$ about $M_1^z = 1$:
\beq
V(1) = -1 +  \left( 1 - M_2^z \right) \left( 1 -M_1^z \right) + O\left( \left( 1 - M_1^z  \right)^2 \right) \ ,
\eeq
which is minimized only at $M_2^z = 1$.  This is in contrast to what happens at precisely $M_1^z = 1$, where any value of $M_2^z$ minimizes the potential at $s=1$. Thus, if $M_1^z$ deviates slightly from 1, which naturally occurs in this Monte Carlo algorithm, there is an energetic bias towards $M_2^z = 1$ \footnote{An analogous noise-induced bias was observed for the problem in Ref.~\cite{q-sig2}}.

\begin{figure}[htbp]
\begin{center}
\subfigure[]{\includegraphics[width=0.48\columnwidth]{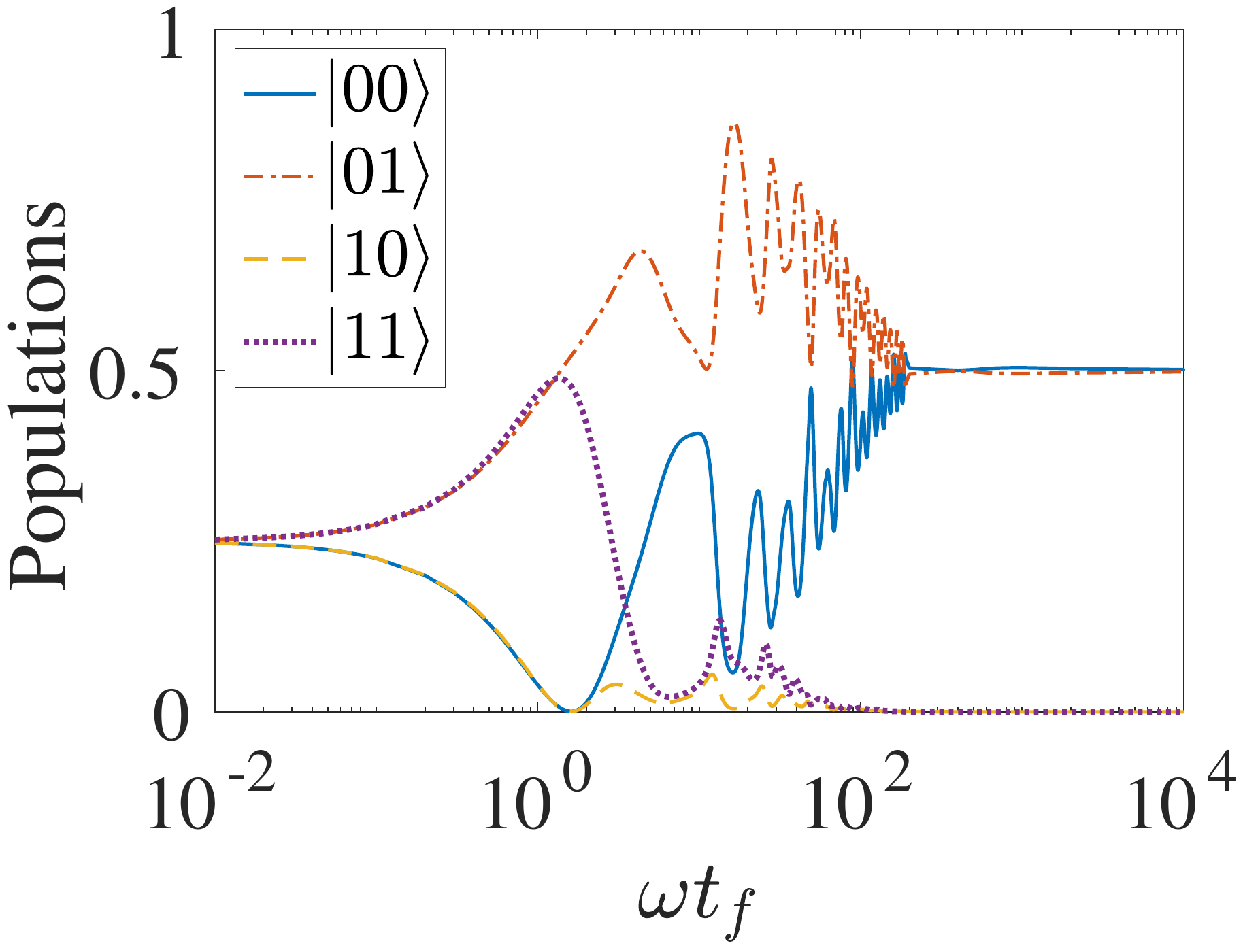} \label{fig:SVD_Pop}}
\subfigure[]{\includegraphics[width=0.48\columnwidth]{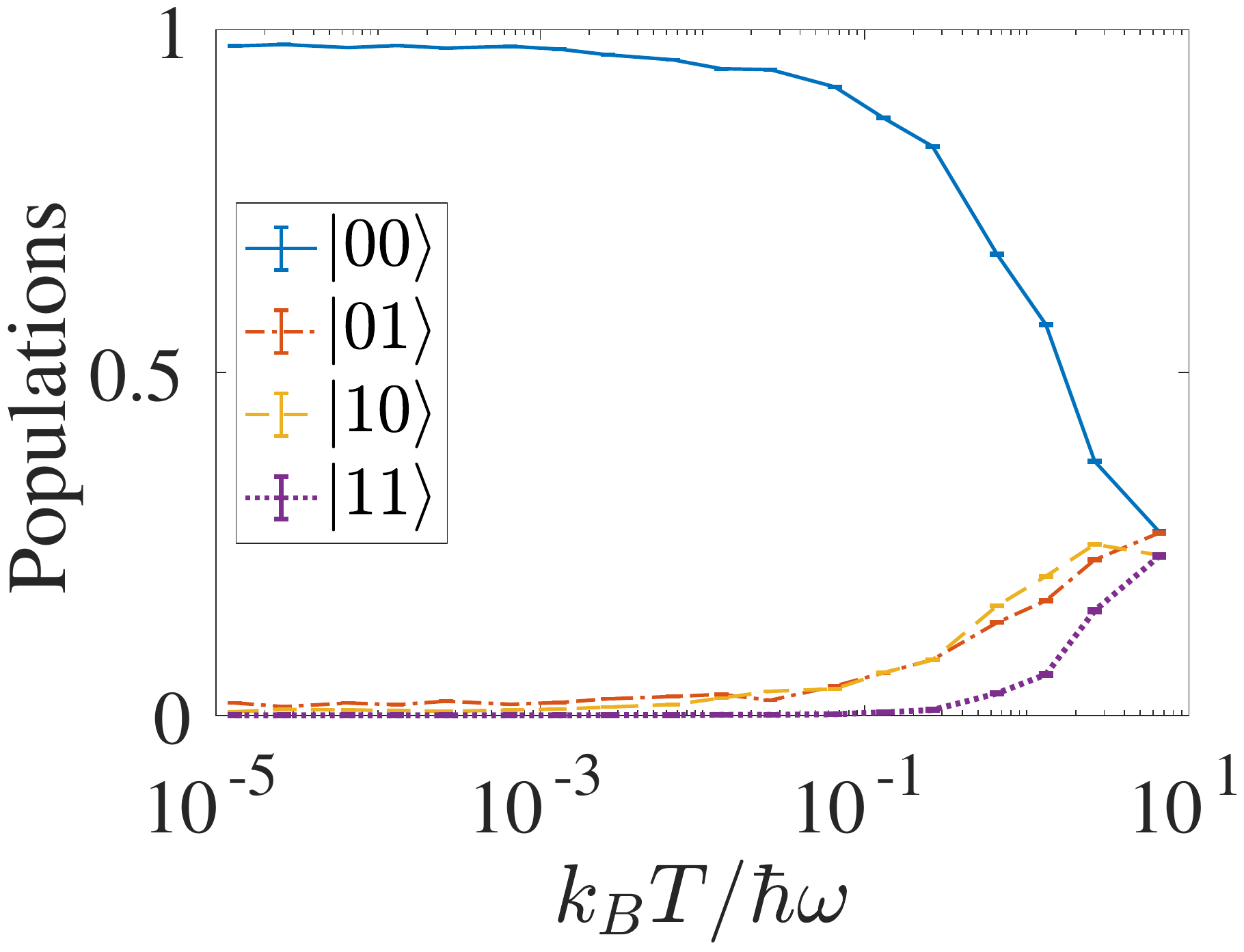} \label{fig:SVMC_Pop}}
   \caption{Computational basis populations for (a)  Spin-vector dynamics  and (b) Spin-vector Monte Carlo with $\alpha = 2$ and $\beta = 0.05$. The results here are equivalent to those of Figs.~\ref{fig:SVD1} and \ref{fig:SVMC_Mz}. For (b), the  simulations use $10^6$ sweeps for each of the $10^4$ independent runs performed.  The error bars correspond to $2\sigma$ confidence intervals generated by $10^3$ bootstraps over the independent runs. } \label{fig:SemiclassicalPopulations}
\end{center}
\end{figure}

In order to more directly compare to the populations of the computational basis states as measured by quantum annealing, we use that the spin vector $\vec{M}_i$ can be interpreted as a vector on the Bloch sphere, so we can assign the probability of measuring a state $|0 \rangle$ for the $i$-th qubit to be $\frac{1}{2} \left( 1 + M_i^z \right)$.  The results in Figs.~\ref{fig:SVD1} and \ref{fig:SVMC_Mz} can then be used to assign probabilities for each of the computational basis states for the 2-qubit system, as shown in Fig.~\ref{fig:SemiclassicalPopulations}.   We see that for the spin-vector dynamics, the long time state corresponding to $\vec{\theta} = (0,\pi/2)$ at $\alpha = 2$ corresponds to equal probabilities of finding the states $\ket{00}$ and $\ket{01}$.  This means that there will always be a finite non-negligible population on the $\ket{00}$ state, in contrast to the results of our unitary dynamics.  Similarly, for Spin-vector Monte Carlo, the strong bias towards $M_1^z = M_2^z =1$ at low temperatures means we have almost all the population on the $\ket{00}$ state.
%

\section{Discussion} \label{sec:Conclusion}
%
 We have proposed a quantum annealing protocol using a two qubit Hamiltonian to validate a tunable antiferromagnetic $XX$ interaction.  The Ising Hamiltonian at the end of the anneal has three ground states, and our experimental signature is the suppression of one of these ground states, the $\ket{00}$ state, for a sufficiently strong antiferromagnetic $XX$ interaction.  Our construction relies on the instantaneous ground state changing character from a symmetric state to an antisymmetric state under the interchange of the two qubits, and the $\ket{00}$ ground state cannot be part of an antisymmetric combination.  
  
While our analysis was done in dimensionless units, it is useful to give a sense of the parameters values for a reasonable choice of energy scale for the Hamiltonian.  If we take $\omega = 1$GHz and $\alpha = 2$, then the closed system evolution shown in Fig.~\ref{fig:OpenSystem1} becomes effectively adiabatic for $t_f \sim 10 \mu s$ with the minimum ground state energy gap of approximately $0.03$GHz at around $s = 0.49$ (see Fig.~\ref{fig:2qGadgetA}). For the coupling to a thermal environment at $12$mK shown in Fig.~\ref{fig:OpenSystem1}, our results suggest that our proposal requires a high level of coherence if we want to see a complete suppression of the $\ket{00}$ state.  However we see from our simulation results of the spin-vector dynamics in Figs.~\ref{fig:SVD} and \ref{fig:SVMC} that the classical analogue fails to exhibit this suppression at the same time scales, so we may not need an absolute suppression signature to clearly distinguish between the quantum dynamics and at least this semiclassical model.

   \begin{figure}[t]
\begin{center}
\subfigure[]{\includegraphics[width=0.48\columnwidth]{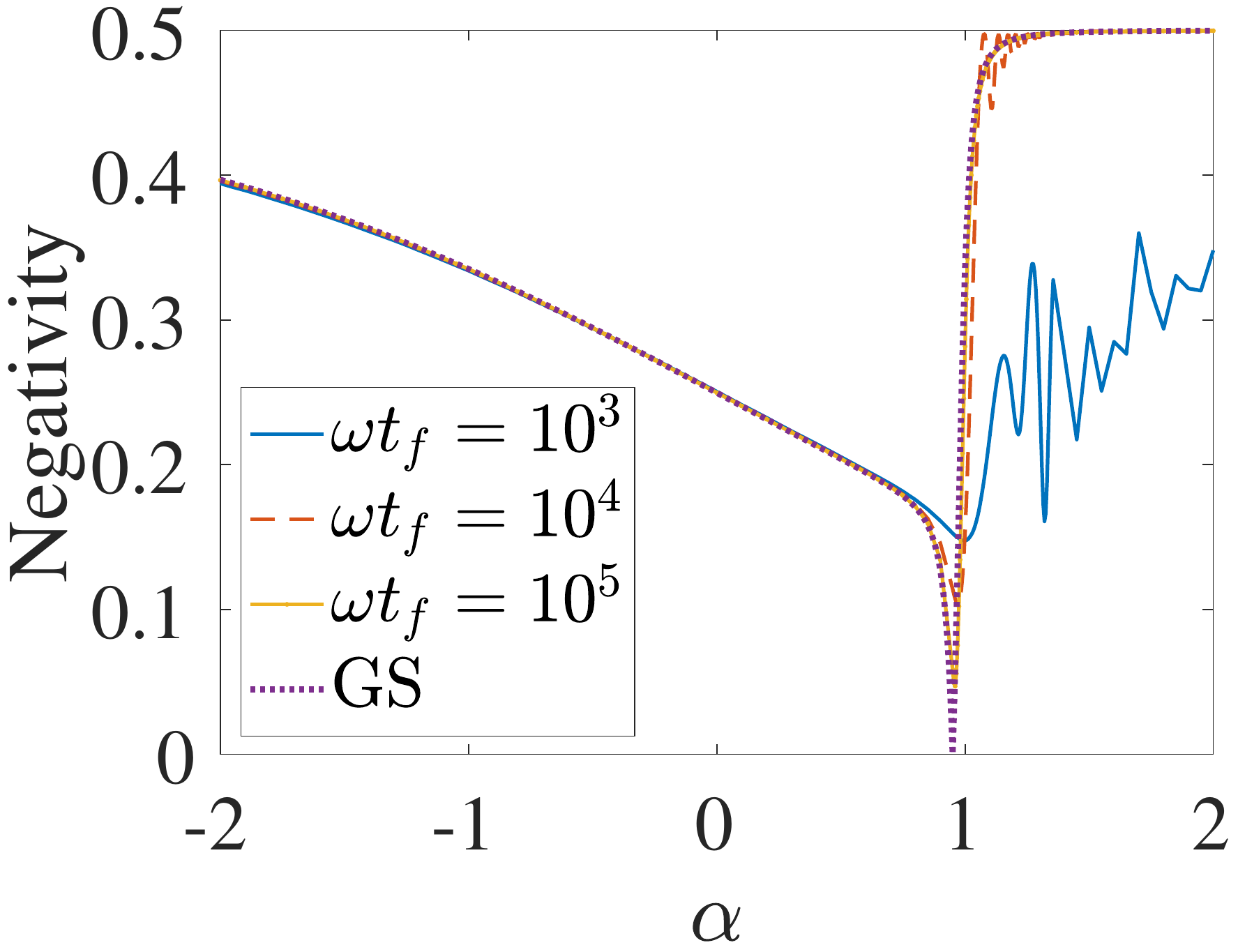}}
\subfigure[]{\includegraphics[width=0.48\columnwidth]{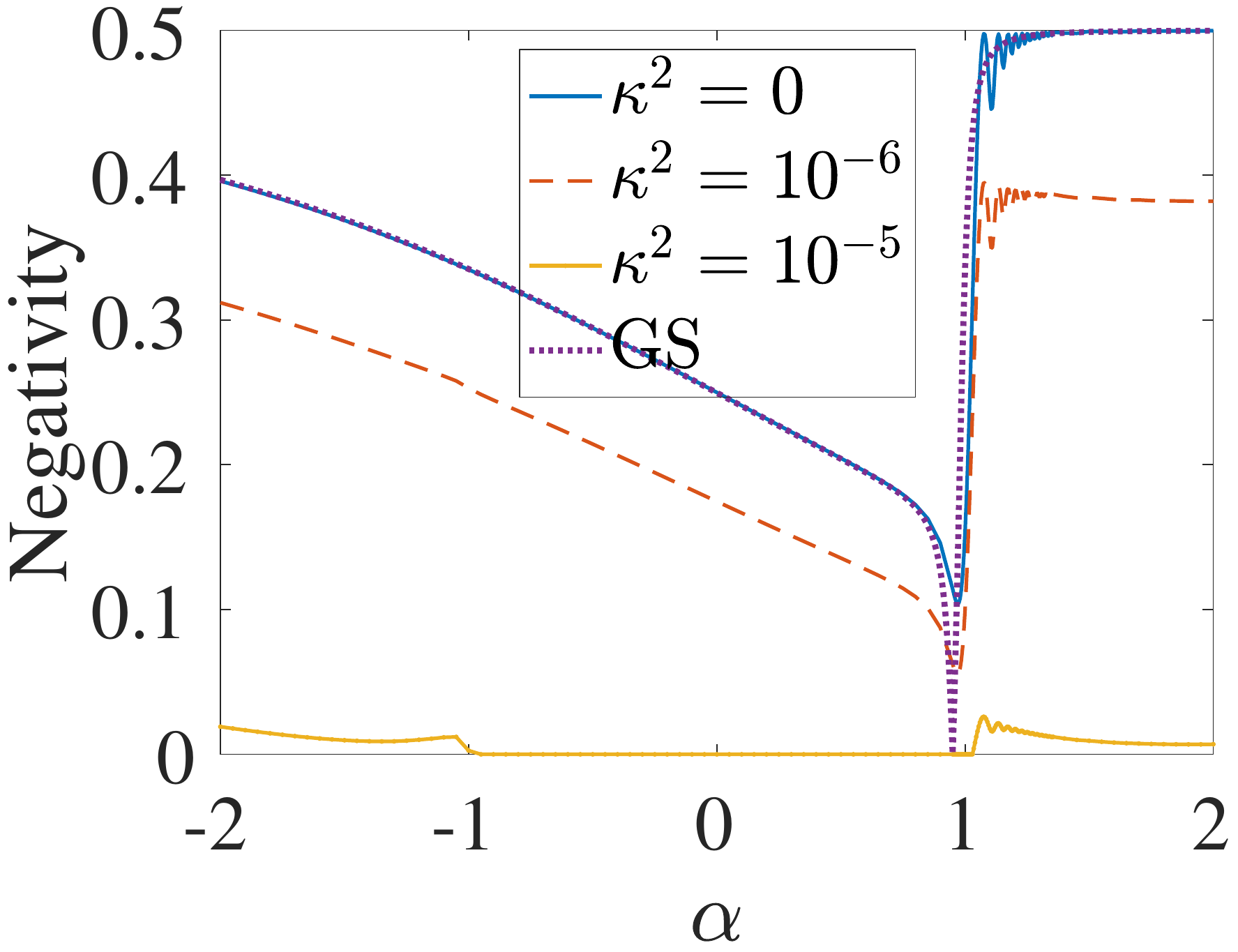}}
   \caption{Negativity at fixed $\beta = 0.05$ for (a) $\kappa^2 = 0$ and varying $t_f$, and (b) $\omega t_f = 10^4$ and varying $\kappa^2$.  Also shown is the ground state value as calculated from perturbation theory (denoted `GS') .}  \label{fig:Negativity}
\end{center}
\end{figure}
 It is interesting to note that the strong change in the character of the ground state as a function of $\alpha$ in Fig.~\ref{fig:2qGadget} is also associated with a large change in the entanglement of the ground state. We characterize this using the negativity \cite{Vidal:02a}, which vanishes for unentangled states and is given by $\mathcal{N}(\rho) = \frac{1}{2} \left( || \rho^{\Gamma} ||_1 - 1 \right)$, where $\rho$ is the density matrix of the system and $\rho^{\Gamma}$ denotes the partial transpose of $\rho$ with respect to one of the qubits and $|| \cdot ||_1$ denotes the trace norm.  We show in Fig.~\ref{fig:Negativity} how the negativity behaves both as a function of annealing time and system-bath coupling, where we observe that close to the ideal case the negativity drops before reaching its maximum value of as we cross $\alpha = 1$.  

We conclude by emphasizing again that the antisymmetric ground state at large $\alpha$ is impossible with only ferromagnetic $XX$ interactions, since it requires negative amplitudes in the ground state that can only be generated by non-stoquastic Hamiltonians.  While the role of such ground states in improving the performance of quantum annealing remains an open research question, the known examples where such an improvement is possible \cite{Nishimori:2016aa,PhysRevA.95.042321,PhysRevA.99.042334} definitely generate ground states with both positive and negative amplitudes \cite{PhysRevA.99.042334}. We hope our proposal will be relevant for testing next generation quantum annealers with such interactions \cite{Kerman_2019}.

\begin{acknowledgments}
We thank Daniel Lidar for useful comments on the manuscript.  Computation for the work described here was supported by the University of Southern California's Center for High-Performance Computing (\url{http://hpcc.usc.edu}) and by ARO grant number W911NF1810227. 
The research is based upon work (partially) supported by the Office of
the Director of National Intelligence (ODNI), Intelligence Advanced
Research Projects Activity (IARPA), via the U.S. Army Research Office
contract W911NF-17-C-0050. The U.S. Government is authorized to reproduce and distribute
reprints for Governmental purposes notwithstanding any copyright annotation thereon.
The views and conclusions contained herein are those of the authors and should not be interpreted as necessarily representing the official policies or endorsements, either expressed or implied, of the ODNI, IARPA, or the U.S. Government. 
\end{acknowledgments}


\begin{thebibliography}{40}%
\makeatletter
\providecommand \@ifxundefined [1]{%
 \@ifx{#1\undefined}
}%
\providecommand \@ifnum [1]{%
 \ifnum #1\expandafter \@firstoftwo
 \else \expandafter \@secondoftwo
 \fi
}%
\providecommand \@ifx [1]{%
 \ifx #1\expandafter \@firstoftwo
 \else \expandafter \@secondoftwo
 \fi
}%
\providecommand \natexlab [1]{#1}%
\providecommand \enquote  [1]{``#1''}%
\providecommand \bibnamefont  [1]{#1}%
\providecommand \bibfnamefont [1]{#1}%
\providecommand \citenamefont [1]{#1}%
\providecommand \href@noop [0]{\@secondoftwo}%
\providecommand \href [0]{\begingroup \@sanitize@url \@href}%
\providecommand \@href[1]{\@@startlink{#1}\@@href}%
\providecommand \@@href[1]{\endgroup#1\@@endlink}%
\providecommand \@sanitize@url [0]{\catcode `\\12\catcode `\$12\catcode
  `\&12\catcode `\#12\catcode `\^12\catcode `\_12\catcode `\%12\relax}%
\providecommand \@@startlink[1]{}%
\providecommand \@@endlink[0]{}%
\providecommand \url  [0]{\begingroup\@sanitize@url \@url }%
\providecommand \@url [1]{\endgroup\@href {#1}{\urlprefix }}%
\providecommand \urlprefix  [0]{URL }%
\providecommand \Eprint [0]{\href }%
\providecommand \doibase [0]{http://dx.doi.org/}%
\providecommand \selectlanguage [0]{\@gobble}%
\providecommand \bibinfo  [0]{\@secondoftwo}%
\providecommand \bibfield  [0]{\@secondoftwo}%
\providecommand \translation [1]{[#1]}%
\providecommand \BibitemOpen [0]{}%
\providecommand \bibitemStop [0]{}%
\providecommand \bibitemNoStop [0]{.\EOS\space}%
\providecommand \EOS [0]{\spacefactor3000\relax}%
\providecommand \BibitemShut  [1]{\csname bibitem#1\endcsname}%
\let\auto@bib@innerbib\@empty
\bibitem [{\citenamefont {Finnila}\ \emph {et~al.}(1994)\citenamefont
  {Finnila}, \citenamefont {Gomez}, \citenamefont {Sebenik}, \citenamefont
  {Stenson},\ and\ \citenamefont {Doll}}]{finnila_quantum_1994}%
  \BibitemOpen
  \bibfield  {author} {\bibinfo {author} {\bibfnamefont {A.~B.}\ \bibnamefont
  {Finnila}}, \bibinfo {author} {\bibfnamefont {M.~A.}\ \bibnamefont {Gomez}},
  \bibinfo {author} {\bibfnamefont {C.}~\bibnamefont {Sebenik}}, \bibinfo
  {author} {\bibfnamefont {C.}~\bibnamefont {Stenson}}, \ and\ \bibinfo
  {author} {\bibfnamefont {J.~D.}\ \bibnamefont {Doll}},\ }\bibfield  {title}
  {\enquote {\bibinfo {title} {Quantum annealing: A new method for minimizing
  multidimensional functions},}\ }\href {\doibase 10.1016/0009-2614(94)00117-0}
  {\bibfield  {journal} {\bibinfo  {journal} {Chemical Physics Letters}\
  }\textbf {\bibinfo {volume} {219}},\ \bibinfo {pages} {343--348} (\bibinfo
  {year} {1994})}\BibitemShut {NoStop}%
\bibitem [{\citenamefont {Brooke}\ \emph {et~al.}(1999)\citenamefont {Brooke},
  \citenamefont {Bitko}, \citenamefont {F.}, \citenamefont {Rosenbaum},\ and\
  \citenamefont {Aeppli}}]{Brooke1999}%
  \BibitemOpen
  \bibfield  {author} {\bibinfo {author} {\bibfnamefont {J.}~\bibnamefont
  {Brooke}}, \bibinfo {author} {\bibfnamefont {D.}~\bibnamefont {Bitko}},
  \bibinfo {author} {\bibfnamefont {T.}~\bibnamefont {F.}}, \bibinfo {author}
  {\bibnamefont {Rosenbaum}}, \ and\ \bibinfo {author} {\bibfnamefont
  {G.}~\bibnamefont {Aeppli}},\ }\bibfield  {title} {\enquote {\bibinfo {title}
  {Quantum annealing of a disordered magnet},}\ }\href {\doibase
  10.1126/science.284.5415.779} {\bibfield  {journal} {\bibinfo  {journal}
  {Science}\ }\textbf {\bibinfo {volume} {284}},\ \bibinfo {pages} {779--781}
  (\bibinfo {year} {1999})}\BibitemShut {NoStop}%
\bibitem [{\citenamefont {Kadowaki}\ and\ \citenamefont
  {Nishimori}(1998)}]{kadowaki_quantum_1998}%
  \BibitemOpen
  \bibfield  {author} {\bibinfo {author} {\bibfnamefont {Tadashi}\ \bibnamefont
  {Kadowaki}}\ and\ \bibinfo {author} {\bibfnamefont {Hidetoshi}\ \bibnamefont
  {Nishimori}},\ }\bibfield  {title} {\enquote {\bibinfo {title} {Quantum
  annealing in the transverse \uppercase{I}sing model},}\ }\href
  {http://journals.aps.org/pre/abstract/10.1103/PhysRevE.58.5355} {\bibfield
  {journal} {\bibinfo  {journal} {Phys. Rev. E}\ }\textbf {\bibinfo {volume}
  {58}},\ \bibinfo {pages} {5355} (\bibinfo {year} {1998})}\BibitemShut
  {NoStop}%
\bibitem [{\citenamefont {Farhi}\ \emph {et~al.}(2001)\citenamefont {Farhi},
  \citenamefont {Goldstone}, \citenamefont {Gutmann}, \citenamefont {Lapan},
  \citenamefont {Lundgren},\ and\ \citenamefont {Preda}}]{Farhi:01}%
  \BibitemOpen
  \bibfield  {author} {\bibinfo {author} {\bibfnamefont {Edward}\ \bibnamefont
  {Farhi}}, \bibinfo {author} {\bibfnamefont {Jeffrey}\ \bibnamefont
  {Goldstone}}, \bibinfo {author} {\bibfnamefont {Sam}\ \bibnamefont
  {Gutmann}}, \bibinfo {author} {\bibfnamefont {Joshua}\ \bibnamefont {Lapan}},
  \bibinfo {author} {\bibfnamefont {Andrew}\ \bibnamefont {Lundgren}}, \ and\
  \bibinfo {author} {\bibfnamefont {Daniel}\ \bibnamefont {Preda}},\ }\bibfield
   {title} {\enquote {\bibinfo {title} {A quantum {Adiabatic} {Evolution}
  {Algorithm} {Applied} to {Random} {Instances} of an {NP}-{Complete}
  {Problem}},}\ }\href {http://www.sciencemag.org/content/292/5516/472}
  {\bibfield  {journal} {\bibinfo  {journal} {Science}\ }\textbf {\bibinfo
  {volume} {292}},\ \bibinfo {pages} {472--475} (\bibinfo {year}
  {2001})}\BibitemShut {NoStop}%
\bibitem [{\citenamefont {Santoro}\ \emph {et~al.}(2002)\citenamefont
  {Santoro}, \citenamefont {Marto\v{n}\'{a}k}, \citenamefont {Tosatti},\ and\
  \citenamefont {Car}}]{Santoro}%
  \BibitemOpen
  \bibfield  {author} {\bibinfo {author} {\bibfnamefont {Giuseppe~E.}\
  \bibnamefont {Santoro}}, \bibinfo {author} {\bibfnamefont {Roman}\
  \bibnamefont {Marto\v{n}\'{a}k}}, \bibinfo {author} {\bibfnamefont {Erio}\
  \bibnamefont {Tosatti}}, \ and\ \bibinfo {author} {\bibfnamefont {Roberto}\
  \bibnamefont {Car}},\ }\bibfield  {title} {\enquote {\bibinfo {title} {Theory
  of quantum annealing of an {I}sing spin glass},}\ }\href
  {http://science.sciencemag.org/content/295/5564/2427} {\bibfield  {journal}
  {\bibinfo  {journal} {Science}\ }\textbf {\bibinfo {volume} {295}},\ \bibinfo
  {pages} {2427--2430} (\bibinfo {year} {2002})}\BibitemShut {NoStop}%
\bibitem [{\citenamefont {Roland}\ and\ \citenamefont
  {Cerf}(2002)}]{Roland:2002ul}%
  \BibitemOpen
  \bibfield  {author} {\bibinfo {author} {\bibfnamefont {J{\'e}r{\'e}mie}\
  \bibnamefont {Roland}}\ and\ \bibinfo {author} {\bibfnamefont {Nicolas~J.}\
  \bibnamefont {Cerf}},\ }\bibfield  {title} {\enquote {\bibinfo {title}
  {Quantum search by local adiabatic evolution},}\ }\href
  {http://link.aps.org/doi/10.1103/PhysRevA.65.042308} {\bibfield  {journal}
  {\bibinfo  {journal} {Phys. Rev. A}\ }\textbf {\bibinfo {volume} {65}},\
  \bibinfo {pages} {042308--} (\bibinfo {year} {2002})}\BibitemShut {NoStop}%
\bibitem [{\citenamefont {Somma}\ \emph {et~al.}(2012)\citenamefont {Somma},
  \citenamefont {Nagaj},\ and\ \citenamefont {Kieferov{\'a}}}]{Somma:2012kx}%
  \BibitemOpen
  \bibfield  {author} {\bibinfo {author} {\bibfnamefont {Rolando~D.}\
  \bibnamefont {Somma}}, \bibinfo {author} {\bibfnamefont {Daniel}\
  \bibnamefont {Nagaj}}, \ and\ \bibinfo {author} {\bibfnamefont {M{\'a}ria}\
  \bibnamefont {Kieferov{\'a}}},\ }\bibfield  {title} {\enquote {\bibinfo
  {title} {Quantum speedup by quantum annealing},}\ }\href
  {http://link.aps.org/doi/10.1103/PhysRevLett.109.050501} {\bibfield
  {journal} {\bibinfo  {journal} {Phys. Rev. Lett.}\ }\textbf {\bibinfo
  {volume} {109}},\ \bibinfo {pages} {050501--} (\bibinfo {year}
  {2012})}\BibitemShut {NoStop}%
\bibitem [{\citenamefont {Bravyi}\ and\ \citenamefont
  {Terhal}(2009)}]{Bravyi:2009sp}%
  \BibitemOpen
  \bibfield  {author} {\bibinfo {author} {\bibfnamefont {S.}~\bibnamefont
  {Bravyi}}\ and\ \bibinfo {author} {\bibfnamefont {B.}~\bibnamefont
  {Terhal}},\ }\bibfield  {title} {\enquote {\bibinfo {title} {Complexity of
  stoquastic frustration-free hamiltonians},}\ }\bibfield  {booktitle} {\emph
  {\bibinfo {booktitle} {SIAM Journal on Computing}},\ }\href {\doibase
  10.1137/08072689X} {\bibfield  {journal} {\bibinfo  {journal} {SIAM Journal
  on Computing}\ }\textbf {\bibinfo {volume} {39}},\ \bibinfo {pages}
  {1462--1485} (\bibinfo {year} {2009})}\BibitemShut {NoStop}%
\bibitem [{\citenamefont {Bravyi}\ \emph {et~al.}(2008)\citenamefont {Bravyi},
  \citenamefont {Divincenzo}, \citenamefont {Oliveira},\ and\ \citenamefont
  {Terhal}}]{Bravyi:QIC08}%
  \BibitemOpen
  \bibfield  {author} {\bibinfo {author} {\bibfnamefont {Sergey}\ \bibnamefont
  {Bravyi}}, \bibinfo {author} {\bibfnamefont {David~P.}\ \bibnamefont
  {Divincenzo}}, \bibinfo {author} {\bibfnamefont {Roberto}\ \bibnamefont
  {Oliveira}}, \ and\ \bibinfo {author} {\bibfnamefont {Barbara~M.}\
  \bibnamefont {Terhal}},\ }\bibfield  {title} {\enquote {\bibinfo {title} {The
  complexity of stoquastic local hamiltonian problems},}\ }\href
  {https://dl.acm.org/doi/10.5555/2011772.2011773} {\bibfield  {journal}
  {\bibinfo  {journal} {Quantum Info. Comput.}\ }\textbf {\bibinfo {volume}
  {8}},\ \bibinfo {pages} {361–385} (\bibinfo {year} {2008})}\BibitemShut
  {NoStop}%
\bibitem [{\citenamefont {Bravyi}\ and\ \citenamefont
  {Hastings}(2017)}]{Bravyi:2014bf}%
  \BibitemOpen
  \bibfield  {author} {\bibinfo {author} {\bibfnamefont {Sergey}\ \bibnamefont
  {Bravyi}}\ and\ \bibinfo {author} {\bibfnamefont {Matthew}\ \bibnamefont
  {Hastings}},\ }\bibfield  {title} {\enquote {\bibinfo {title} {On complexity
  of the quantum ising model},}\ }\href {\doibase 10.1007/s00220-016-2787-4}
  {\bibfield  {journal} {\bibinfo  {journal} {Communications in Mathematical
  Physics}\ }\textbf {\bibinfo {volume} {349}},\ \bibinfo {pages} {1--45}
  (\bibinfo {year} {2017})}\BibitemShut {NoStop}%
\bibitem [{\citenamefont {Marvian}\ \emph {et~al.}(2019)\citenamefont
  {Marvian}, \citenamefont {Lidar},\ and\ \citenamefont {Hen}}]{Marvian2019}%
  \BibitemOpen
  \bibfield  {author} {\bibinfo {author} {\bibfnamefont {Milad}\ \bibnamefont
  {Marvian}}, \bibinfo {author} {\bibfnamefont {Daniel~A.}\ \bibnamefont
  {Lidar}}, \ and\ \bibinfo {author} {\bibfnamefont {Itay}\ \bibnamefont
  {Hen}},\ }\bibfield  {title} {\enquote {\bibinfo {title} {On the
  computational complexity of curing non-stoquastic hamiltonians},}\ }\href
  {\doibase 10.1038/s41467-019-09501-6} {\bibfield  {journal} {\bibinfo
  {journal} {Nature Communications}\ }\textbf {\bibinfo {volume} {10}},\
  \bibinfo {pages} {1571} (\bibinfo {year} {2019})}\BibitemShut {NoStop}%
\bibitem [{\citenamefont {Klassen}\ and\ \citenamefont
  {Terhal}(2019)}]{Klassen2019twolocalqubit}%
  \BibitemOpen
  \bibfield  {author} {\bibinfo {author} {\bibfnamefont {Joel}\ \bibnamefont
  {Klassen}}\ and\ \bibinfo {author} {\bibfnamefont {Barbara~M.}\ \bibnamefont
  {Terhal}},\ }\bibfield  {title} {\enquote {\bibinfo {title} {Two-local qubit
  {H}amiltonians: when are they stoquastic?}}\ }\href {\doibase
  10.22331/q-2019-05-06-139} {\bibfield  {journal} {\bibinfo  {journal}
  {{Quantum}}\ }\textbf {\bibinfo {volume} {3}},\ \bibinfo {pages} {139}
  (\bibinfo {year} {2019})}\BibitemShut {NoStop}%
\bibitem [{\citenamefont {{Klassen}}\ \emph {et~al.}(2019)\citenamefont
  {{Klassen}}, \citenamefont {{Marvian}}, \citenamefont {{Piddock}},
  \citenamefont {{Ioannou}}, \citenamefont {{Hen}},\ and\ \citenamefont
  {{Terhal}}}]{Klassen2019b}%
  \BibitemOpen
  \bibfield  {author} {\bibinfo {author} {\bibfnamefont {Joel}\ \bibnamefont
  {{Klassen}}}, \bibinfo {author} {\bibfnamefont {Milad}\ \bibnamefont
  {{Marvian}}}, \bibinfo {author} {\bibfnamefont {Stephen}\ \bibnamefont
  {{Piddock}}}, \bibinfo {author} {\bibfnamefont {Marios}\ \bibnamefont
  {{Ioannou}}}, \bibinfo {author} {\bibfnamefont {Itay}\ \bibnamefont {{Hen}}},
  \ and\ \bibinfo {author} {\bibfnamefont {Barbara}\ \bibnamefont {{Terhal}}},\
  }\bibfield  {title} {\enquote {\bibinfo {title} {{Hardness and Ease of Curing
  the Sign Problem for Two-Local Qubit Hamiltonians}},}\ }\href@noop {}
  {\bibfield  {journal} {\bibinfo  {journal} {arXiv e-prints}\ ,\ \bibinfo
  {eid} {arXiv:1906.08800}} (\bibinfo {year} {2019})},\ \Eprint
  {http://arxiv.org/abs/1906.08800} {arXiv:1906.08800 [quant-ph]} \BibitemShut
  {NoStop}%
\bibitem [{\citenamefont {Farhi}\ \emph {et~al.}(2000)\citenamefont {Farhi},
  \citenamefont {Goldstone}, \citenamefont {Gutmann},\ and\ \citenamefont
  {Sipser}}]{Farhi:00}%
  \BibitemOpen
  \bibfield  {author} {\bibinfo {author} {\bibfnamefont {Edward}\ \bibnamefont
  {Farhi}}, \bibinfo {author} {\bibfnamefont {Jeffrey}\ \bibnamefont
  {Goldstone}}, \bibinfo {author} {\bibfnamefont {Sam}\ \bibnamefont
  {Gutmann}}, \ and\ \bibinfo {author} {\bibfnamefont {Michael}\ \bibnamefont
  {Sipser}},\ }\bibfield  {title} {\enquote {\bibinfo {title} {Quantum
  {Computation} by {Adiabatic} {Evolution}},}\ }\href
  {http://arxiv.org/abs/quant-ph/0001106} {\bibfield  {journal} {\bibinfo
  {journal} {arXiv:quant-ph/0001106}\ } (\bibinfo {year} {2000})}\BibitemShut
  {NoStop}%
\bibitem [{\citenamefont {Isakov}\ \emph {et~al.}(2016)\citenamefont {Isakov},
  \citenamefont {Mazzola}, \citenamefont {Smelyanskiy}, \citenamefont {Jiang},
  \citenamefont {Boixo}, \citenamefont {Neven},\ and\ \citenamefont
  {Troyer}}]{2015arXiv151008057I}%
  \BibitemOpen
  \bibfield  {author} {\bibinfo {author} {\bibfnamefont {Sergei~V.}\
  \bibnamefont {Isakov}}, \bibinfo {author} {\bibfnamefont {Guglielmo}\
  \bibnamefont {Mazzola}}, \bibinfo {author} {\bibfnamefont {Vadim~N.}\
  \bibnamefont {Smelyanskiy}}, \bibinfo {author} {\bibfnamefont {Zhang}\
  \bibnamefont {Jiang}}, \bibinfo {author} {\bibfnamefont {Sergio}\
  \bibnamefont {Boixo}}, \bibinfo {author} {\bibfnamefont {Hartmut}\
  \bibnamefont {Neven}}, \ and\ \bibinfo {author} {\bibfnamefont {Matthias}\
  \bibnamefont {Troyer}},\ }\bibfield  {title} {\enquote {\bibinfo {title}
  {{Understanding Quantum Tunneling through Quantum Monte Carlo
  Simulations}},}\ }\href
  {http://link.aps.org/doi/10.1103/PhysRevLett.117.180402} {\bibfield
  {journal} {\bibinfo  {journal} {Physical Review Letters}\ }\textbf {\bibinfo
  {volume} {117}},\ \bibinfo {pages} {180402--} (\bibinfo {year}
  {2016})}\BibitemShut {NoStop}%
\bibitem [{\citenamefont {Hastings}(2013)}]{Hastings:2013kk}%
  \BibitemOpen
  \bibfield  {author} {\bibinfo {author} {\bibfnamefont {Matthew~B.}\
  \bibnamefont {Hastings}},\ }\bibfield  {title} {\enquote {\bibinfo {title}
  {Obstructions to classically simulating the quantum adiabatic algorithm},}\
  }\href {https://dl.acm.org/doi/10.5555/2535639.2535647} {\bibfield  {journal}
  {\bibinfo  {journal} {Quantum Info. Comput.}\ }\textbf {\bibinfo {volume}
  {13}},\ \bibinfo {pages} {1038–1076} (\bibinfo {year} {2013})}\BibitemShut
  {NoStop}%
\bibitem [{\citenamefont {Andriyash}\ and\ \citenamefont
  {Amin}(2017)}]{Andriyash:2017aa}%
  \BibitemOpen
  \bibfield  {author} {\bibinfo {author} {\bibfnamefont {Evgeny}\ \bibnamefont
  {Andriyash}}\ and\ \bibinfo {author} {\bibfnamefont {Mohammad~H.}\
  \bibnamefont {Amin}},\ }\bibfield  {title} {\enquote {\bibinfo {title} {{Can
  quantum Monte Carlo simulate quantum annealing?}}}\ }\href
  {http://arXiv.org/abs/1703.09277} {\bibfield  {journal} {\bibinfo  {journal}
  {arXiv:1703.09277}\ } (\bibinfo {year} {2017})}\BibitemShut {NoStop}%
\bibitem [{\citenamefont {Troyer}\ and\ \citenamefont
  {Wiese}(2005)}]{PhysRevLett.94.170201}%
  \BibitemOpen
  \bibfield  {author} {\bibinfo {author} {\bibfnamefont {Matthias}\
  \bibnamefont {Troyer}}\ and\ \bibinfo {author} {\bibfnamefont {Uwe-Jens}\
  \bibnamefont {Wiese}},\ }\bibfield  {title} {\enquote {\bibinfo {title}
  {Computational complexity and fundamental limitations to fermionic quantum
  monte carlo simulations},}\ }\href {\doibase 10.1103/PhysRevLett.94.170201}
  {\bibfield  {journal} {\bibinfo  {journal} {Phys. Rev. Lett.}\ }\textbf
  {\bibinfo {volume} {94}},\ \bibinfo {pages} {170201} (\bibinfo {year}
  {2005})}\BibitemShut {NoStop}%
\bibitem [{\citenamefont {Susa}\ \emph {et~al.}(2017)\citenamefont {Susa},
  \citenamefont {Jadebeck},\ and\ \citenamefont
  {Nishimori}}]{PhysRevA.95.042321}%
  \BibitemOpen
  \bibfield  {author} {\bibinfo {author} {\bibfnamefont {Yuki}\ \bibnamefont
  {Susa}}, \bibinfo {author} {\bibfnamefont {Johann~F.}\ \bibnamefont
  {Jadebeck}}, \ and\ \bibinfo {author} {\bibfnamefont {Hidetoshi}\
  \bibnamefont {Nishimori}},\ }\bibfield  {title} {\enquote {\bibinfo {title}
  {Relation between quantum fluctuations and the performance enhancement of
  quantum annealing in a nonstoquastic hamiltonian},}\ }\href {\doibase
  10.1103/PhysRevA.95.042321} {\bibfield  {journal} {\bibinfo  {journal} {Phys.
  Rev. A}\ }\textbf {\bibinfo {volume} {95}},\ \bibinfo {pages} {042321}
  (\bibinfo {year} {2017})}\BibitemShut {NoStop}%
\bibitem [{\citenamefont {Nishimori}\ and\ \citenamefont
  {Takada}(2017)}]{Nishimori:2016aa}%
  \BibitemOpen
  \bibfield  {author} {\bibinfo {author} {\bibfnamefont {Hidetoshi}\
  \bibnamefont {Nishimori}}\ and\ \bibinfo {author} {\bibfnamefont {Kabuki}\
  \bibnamefont {Takada}},\ }\bibfield  {title} {\enquote {\bibinfo {title}
  {Exponential enhancement of the efficiency of quantum annealing by
  non-stoquastic hamiltonians},}\ }\href {\doibase 10.3389/fict.2017.00002}
  {\bibfield  {journal} {\bibinfo  {journal} {Frontiers in ICT}\ }\textbf
  {\bibinfo {volume} {4}},\ \bibinfo {pages} {2} (\bibinfo {year}
  {2017})}\BibitemShut {NoStop}%
\bibitem [{\citenamefont {Crosson}\ \emph {et~al.}(2014)\citenamefont
  {Crosson}, \citenamefont {Farhi}, \citenamefont {Lin}, \citenamefont {Lin},\
  and\ \citenamefont {Shor}}]{crosson2014different}%
  \BibitemOpen
  \bibfield  {author} {\bibinfo {author} {\bibfnamefont {Elizabeth}\
  \bibnamefont {Crosson}}, \bibinfo {author} {\bibfnamefont {Edward}\
  \bibnamefont {Farhi}}, \bibinfo {author} {\bibfnamefont {Cedric Yen-Yu}\
  \bibnamefont {Lin}}, \bibinfo {author} {\bibfnamefont {Han-Hsuan}\
  \bibnamefont {Lin}}, \ and\ \bibinfo {author} {\bibfnamefont {Peter}\
  \bibnamefont {Shor}},\ }\bibfield  {title} {\enquote {\bibinfo {title}
  {Different strategies for optimization using the quantum adiabatic
  algorithm},}\ }\href {http://arxiv.org/abs/1401.7320} {\bibfield  {journal}
  {\bibinfo  {journal} {arXiv preprint arXiv:1401.7320}\ } (\bibinfo {year}
  {2014})}\BibitemShut {NoStop}%
\bibitem [{\citenamefont {Hormozi}\ \emph {et~al.}(2017)\citenamefont
  {Hormozi}, \citenamefont {Brown}, \citenamefont {Carleo},\ and\ \citenamefont
  {Troyer}}]{Hormozi:2016aa}%
  \BibitemOpen
  \bibfield  {author} {\bibinfo {author} {\bibfnamefont {Layla}\ \bibnamefont
  {Hormozi}}, \bibinfo {author} {\bibfnamefont {Ethan~W.}\ \bibnamefont
  {Brown}}, \bibinfo {author} {\bibfnamefont {Giuseppe}\ \bibnamefont
  {Carleo}}, \ and\ \bibinfo {author} {\bibfnamefont {Matthias}\ \bibnamefont
  {Troyer}},\ }\bibfield  {title} {\enquote {\bibinfo {title} {Nonstoquastic
  hamiltonians and quantum annealing of an ising spin glass},}\ }\href
  {\doibase 10.1103/PhysRevB.95.184416} {\bibfield  {journal} {\bibinfo
  {journal} {Phys. Rev. B}\ }\textbf {\bibinfo {volume} {95}},\ \bibinfo
  {pages} {184416} (\bibinfo {year} {2017})}\BibitemShut {NoStop}%
\bibitem [{\citenamefont {Albash}(2019)}]{PhysRevA.99.042334}%
  \BibitemOpen
  \bibfield  {author} {\bibinfo {author} {\bibfnamefont {Tameem}\ \bibnamefont
  {Albash}},\ }\bibfield  {title} {\enquote {\bibinfo {title} {Role of
  nonstoquastic catalysts in quantum adiabatic optimization},}\ }\href
  {\doibase 10.1103/PhysRevA.99.042334} {\bibfield  {journal} {\bibinfo
  {journal} {Phys. Rev. A}\ }\textbf {\bibinfo {volume} {99}},\ \bibinfo
  {pages} {042334} (\bibinfo {year} {2019})}\BibitemShut {NoStop}%
\bibitem [{Note1()}]{Note1}%
  \BibitemOpen
  \bibinfo {note} {We note that tunable non-stoquastic interactions are
  necessary for universal adiabatic quantum computing \cite {Biamonte:07},
  although we are not considering this question here.}\BibitemShut {Stop}%
\bibitem [{\citenamefont {{Ozfidan}}\ \emph {et~al.}(2019)\citenamefont
  {{Ozfidan}}, \citenamefont {{Deng}}, \citenamefont {{Smirnov}}, \citenamefont
  {{Lanting}}, \citenamefont {{Harris}}, \citenamefont {{Swenson}},
  \citenamefont {{Whittaker}}, \citenamefont {{Altomare}}, \citenamefont
  {{Babcock}}, \citenamefont {{Baron}}, \citenamefont {{Berkley}},
  \citenamefont {{Boothby}}, \citenamefont {{Christiani}}, \citenamefont
  {{Bunyk}}, \citenamefont {{Enderud}}, \citenamefont {{Evert}}, \citenamefont
  {{Hager}}, \citenamefont {{Hajda}}, \citenamefont {{Hilton}}, \citenamefont
  {{Huang}}, \citenamefont {{Hoskinson}}, \citenamefont {{Johnson}},
  \citenamefont {{Jooya}}, \citenamefont {{Ladizinsky}}, \citenamefont
  {{Ladizinsky}}, \citenamefont {{Li}}, \citenamefont {{MacDonald}},
  \citenamefont {{Marsden}}, \citenamefont {{Marsden}}, \citenamefont
  {{Medina}}, \citenamefont {{Molavi}}, \citenamefont {{Neufeld}},
  \citenamefont {{Nissen}}, \citenamefont {{Norouzpour}}, \citenamefont {{Oh}},
  \citenamefont {{Pavlov}}, \citenamefont {{Perminov}}, \citenamefont
  {{Poulin-Lamarre}}, \citenamefont {{Reis}}, \citenamefont {{Prescott}},
  \citenamefont {{Rich}}, \citenamefont {{Sato}}, \citenamefont {{Sterling}},
  \citenamefont {{Tsai}}, \citenamefont {{Volkmann}}, \citenamefont
  {{Wilkinson}}, \citenamefont {{Yao}},\ and\ \citenamefont
  {{Amin}}}]{DWaveNonStoq}%
  \BibitemOpen
  \bibfield  {author} {\bibinfo {author} {\bibfnamefont {I.}~\bibnamefont
  {{Ozfidan}}}, \bibinfo {author} {\bibfnamefont {C.}~\bibnamefont {{Deng}}},
  \bibinfo {author} {\bibfnamefont {A.~Y.}\ \bibnamefont {{Smirnov}}}, \bibinfo
  {author} {\bibfnamefont {T.}~\bibnamefont {{Lanting}}}, \bibinfo {author}
  {\bibfnamefont {R.}~\bibnamefont {{Harris}}}, \bibinfo {author}
  {\bibfnamefont {L.}~\bibnamefont {{Swenson}}}, \bibinfo {author}
  {\bibfnamefont {J.}~\bibnamefont {{Whittaker}}}, \bibinfo {author}
  {\bibfnamefont {F.}~\bibnamefont {{Altomare}}}, \bibinfo {author}
  {\bibfnamefont {M.}~\bibnamefont {{Babcock}}}, \bibinfo {author}
  {\bibfnamefont {C.}~\bibnamefont {{Baron}}}, \bibinfo {author} {\bibfnamefont
  {A.~J.}\ \bibnamefont {{Berkley}}}, \bibinfo {author} {\bibfnamefont
  {K.}~\bibnamefont {{Boothby}}}, \bibinfo {author} {\bibfnamefont
  {H.}~\bibnamefont {{Christiani}}}, \bibinfo {author} {\bibfnamefont
  {P.}~\bibnamefont {{Bunyk}}}, \bibinfo {author} {\bibfnamefont
  {C.}~\bibnamefont {{Enderud}}}, \bibinfo {author} {\bibfnamefont
  {B.}~\bibnamefont {{Evert}}}, \bibinfo {author} {\bibfnamefont
  {M.}~\bibnamefont {{Hager}}}, \bibinfo {author} {\bibfnamefont
  {A.}~\bibnamefont {{Hajda}}}, \bibinfo {author} {\bibfnamefont
  {J.}~\bibnamefont {{Hilton}}}, \bibinfo {author} {\bibfnamefont
  {S.}~\bibnamefont {{Huang}}}, \bibinfo {author} {\bibfnamefont
  {E.}~\bibnamefont {{Hoskinson}}}, \bibinfo {author} {\bibfnamefont {M.~W.}\
  \bibnamefont {{Johnson}}}, \bibinfo {author} {\bibfnamefont {K.}~\bibnamefont
  {{Jooya}}}, \bibinfo {author} {\bibfnamefont {E.}~\bibnamefont
  {{Ladizinsky}}}, \bibinfo {author} {\bibfnamefont {N.}~\bibnamefont
  {{Ladizinsky}}}, \bibinfo {author} {\bibfnamefont {R.}~\bibnamefont {{Li}}},
  \bibinfo {author} {\bibfnamefont {A.}~\bibnamefont {{MacDonald}}}, \bibinfo
  {author} {\bibfnamefont {D.}~\bibnamefont {{Marsden}}}, \bibinfo {author}
  {\bibfnamefont {G.}~\bibnamefont {{Marsden}}}, \bibinfo {author}
  {\bibfnamefont {T.}~\bibnamefont {{Medina}}}, \bibinfo {author}
  {\bibfnamefont {R.}~\bibnamefont {{Molavi}}}, \bibinfo {author}
  {\bibfnamefont {R.}~\bibnamefont {{Neufeld}}}, \bibinfo {author}
  {\bibfnamefont {M.}~\bibnamefont {{Nissen}}}, \bibinfo {author}
  {\bibfnamefont {M.}~\bibnamefont {{Norouzpour}}}, \bibinfo {author}
  {\bibfnamefont {T.}~\bibnamefont {{Oh}}}, \bibinfo {author} {\bibfnamefont
  {I.}~\bibnamefont {{Pavlov}}}, \bibinfo {author} {\bibfnamefont
  {I.}~\bibnamefont {{Perminov}}}, \bibinfo {author} {\bibfnamefont
  {G.}~\bibnamefont {{Poulin-Lamarre}}}, \bibinfo {author} {\bibfnamefont
  {M.}~\bibnamefont {{Reis}}}, \bibinfo {author} {\bibfnamefont
  {T.}~\bibnamefont {{Prescott}}}, \bibinfo {author} {\bibfnamefont
  {C.}~\bibnamefont {{Rich}}}, \bibinfo {author} {\bibfnamefont
  {Y.}~\bibnamefont {{Sato}}}, \bibinfo {author} {\bibfnamefont
  {G.}~\bibnamefont {{Sterling}}}, \bibinfo {author} {\bibfnamefont
  {N.}~\bibnamefont {{Tsai}}}, \bibinfo {author} {\bibfnamefont
  {M.}~\bibnamefont {{Volkmann}}}, \bibinfo {author} {\bibfnamefont
  {W.}~\bibnamefont {{Wilkinson}}}, \bibinfo {author} {\bibfnamefont
  {J.}~\bibnamefont {{Yao}}}, \ and\ \bibinfo {author} {\bibfnamefont {M.~H.}\
  \bibnamefont {{Amin}}},\ }\bibfield  {title} {\enquote {\bibinfo {title}
  {{Demonstration of nonstoquastic Hamiltonian in coupled superconducting flux
  qubits}},}\ }\href@noop {} {\bibfield  {journal} {\bibinfo  {journal} {arXiv
  e-prints}\ ,\ \bibinfo {eid} {arXiv:1903.06139}} (\bibinfo {year} {2019})},\
  \Eprint {http://arxiv.org/abs/1903.06139} {arXiv:1903.06139 [quant-ph]}
  \BibitemShut {NoStop}%
\bibitem [{\citenamefont {Kerman}(2019)}]{Kerman_2019}%
  \BibitemOpen
  \bibfield  {author} {\bibinfo {author} {\bibfnamefont {Andrew~J}\
  \bibnamefont {Kerman}},\ }\bibfield  {title} {\enquote {\bibinfo {title}
  {Superconducting qubit circuit emulation of a vector spin-1/2},}\ }\href
  {\doibase 10.1088/1367-2630/ab2ee7} {\bibfield  {journal} {\bibinfo
  {journal} {New Journal of Physics}\ }\textbf {\bibinfo {volume} {21}},\
  \bibinfo {pages} {073030} (\bibinfo {year} {2019})}\BibitemShut {NoStop}%
\bibitem [{\citenamefont {Albash}\ \emph {et~al.}(2012)\citenamefont {Albash},
  \citenamefont {Boixo}, \citenamefont {Lidar},\ and\ \citenamefont
  {Zanardi}}]{ABLZ:12-SI}%
  \BibitemOpen
  \bibfield  {author} {\bibinfo {author} {\bibfnamefont {Tameem}\ \bibnamefont
  {Albash}}, \bibinfo {author} {\bibfnamefont {Sergio}\ \bibnamefont {Boixo}},
  \bibinfo {author} {\bibfnamefont {Daniel~A}\ \bibnamefont {Lidar}}, \ and\
  \bibinfo {author} {\bibfnamefont {Paolo}\ \bibnamefont {Zanardi}},\
  }\bibfield  {title} {\enquote {\bibinfo {title} {{Quantum adiabatic Markovian
  master equations}},}\ }\href {\doibase 10.1088/1367-2630/14/12/123016}
  {\bibfield  {journal} {\bibinfo  {journal} {New J. of Phys.}\ }\textbf
  {\bibinfo {volume} {14}},\ \bibinfo {pages} {123016} (\bibinfo {year}
  {2012})}\BibitemShut {NoStop}%
\bibitem [{\citenamefont {Venuti}\ \emph {et~al.}(2016)\citenamefont {Venuti},
  \citenamefont {Albash}, \citenamefont {Lidar},\ and\ \citenamefont
  {Zanardi}}]{PhysRevA.93.032118}%
  \BibitemOpen
  \bibfield  {author} {\bibinfo {author} {\bibfnamefont {Lorenzo~Campos}\
  \bibnamefont {Venuti}}, \bibinfo {author} {\bibfnamefont {Tameem}\
  \bibnamefont {Albash}}, \bibinfo {author} {\bibfnamefont {Daniel~A.}\
  \bibnamefont {Lidar}}, \ and\ \bibinfo {author} {\bibfnamefont {Paolo}\
  \bibnamefont {Zanardi}},\ }\bibfield  {title} {\enquote {\bibinfo {title}
  {Adiabaticity in open quantum systems},}\ }\href {\doibase
  10.1103/PhysRevA.93.032118} {\bibfield  {journal} {\bibinfo  {journal} {Phys.
  Rev. A}\ }\textbf {\bibinfo {volume} {93}},\ \bibinfo {pages} {032118}
  (\bibinfo {year} {2016})}\BibitemShut {NoStop}%
\bibitem [{\citenamefont {Klauder}(1979)}]{klauder1979path}%
  \BibitemOpen
  \bibfield  {author} {\bibinfo {author} {\bibfnamefont {John~R}\ \bibnamefont
  {Klauder}},\ }\bibfield  {title} {\enquote {\bibinfo {title} {Path integrals
  and stationary-phase approximations},}\ }\href
  {http://journals.aps.org/prd/abstract/10.1103/PhysRevD.19.2349} {\bibfield
  {journal} {\bibinfo  {journal} {Phys. Rev. D}\ }\textbf {\bibinfo {volume}
  {19}},\ \bibinfo {pages} {2349} (\bibinfo {year} {1979})}\BibitemShut
  {NoStop}%
\bibitem [{\citenamefont {Radcliffe}(1971)}]{Radcliffe_1971}%
  \BibitemOpen
  \bibfield  {author} {\bibinfo {author} {\bibfnamefont {J~M}\ \bibnamefont
  {Radcliffe}},\ }\bibfield  {title} {\enquote {\bibinfo {title} {Some
  properties of coherent spin states},}\ }\href {\doibase
  10.1088/0305-4470/4/3/009} {\bibfield  {journal} {\bibinfo  {journal}
  {Journal of Physics A: General Physics}\ }\textbf {\bibinfo {volume} {4}},\
  \bibinfo {pages} {313--323} (\bibinfo {year} {1971})}\BibitemShut {NoStop}%
\bibitem [{\citenamefont {Owerre}\ and\ \citenamefont
  {Paranjape}(2015)}]{owerre2015macroscopic}%
  \BibitemOpen
  \bibfield  {author} {\bibinfo {author} {\bibfnamefont {SA}~\bibnamefont
  {Owerre}}\ and\ \bibinfo {author} {\bibfnamefont {MB}~\bibnamefont
  {Paranjape}},\ }\bibfield  {title} {\enquote {\bibinfo {title} {Macroscopic
  quantum tunneling and quantum--classical phase transitions of the escape rate
  in large spin systems},}\ }\href
  {http://www.sciencedirect.com/science/article/pii/S0370157314003147}
  {\bibfield  {journal} {\bibinfo  {journal} {Physics Reports}\ }\textbf
  {\bibinfo {volume} {546}},\ \bibinfo {pages} {1--60} (\bibinfo {year}
  {2015})}\BibitemShut {NoStop}%
\bibitem [{\citenamefont {Albash}\ \emph {et~al.}(2015)\citenamefont {Albash},
  \citenamefont {Vinci}, \citenamefont {Mishra}, \citenamefont {Warburton},\
  and\ \citenamefont {Lidar}}]{q-sig2}%
  \BibitemOpen
  \bibfield  {author} {\bibinfo {author} {\bibfnamefont {Tameem}\ \bibnamefont
  {Albash}}, \bibinfo {author} {\bibfnamefont {Walter}\ \bibnamefont {Vinci}},
  \bibinfo {author} {\bibfnamefont {Anurag}\ \bibnamefont {Mishra}}, \bibinfo
  {author} {\bibfnamefont {Paul~A.}\ \bibnamefont {Warburton}}, \ and\ \bibinfo
  {author} {\bibfnamefont {Daniel~A.}\ \bibnamefont {Lidar}},\ }\bibfield
  {title} {\enquote {\bibinfo {title} {Consistency tests of classical and
  quantum models for a quantum annealer},}\ }\href
  {http://link.aps.org/doi/10.1103/PhysRevA.91.042314} {\bibfield  {journal}
  {\bibinfo  {journal} {Phys. Rev. A}\ }\textbf {\bibinfo {volume} {91}},\
  \bibinfo {pages} {042314--} (\bibinfo {year} {2015})}\BibitemShut {NoStop}%
\bibitem [{\citenamefont {Metropolis}\ \emph {et~al.}(1953)\citenamefont
  {Metropolis}, \citenamefont {Rosenbluth}, \citenamefont {Rosenbluth},
  \citenamefont {Teller},\ and\ \citenamefont {Teller}}]{Metropolis1953}%
  \BibitemOpen
  \bibfield  {author} {\bibinfo {author} {\bibfnamefont {Nicholas}\
  \bibnamefont {Metropolis}}, \bibinfo {author} {\bibfnamefont {Arianna~W.}\
  \bibnamefont {Rosenbluth}}, \bibinfo {author} {\bibfnamefont {Marshall~N.}\
  \bibnamefont {Rosenbluth}}, \bibinfo {author} {\bibfnamefont {Augusta~H.}\
  \bibnamefont {Teller}}, \ and\ \bibinfo {author} {\bibfnamefont {Edward}\
  \bibnamefont {Teller}},\ }\bibfield  {title} {\enquote {\bibinfo {title}
  {Equation of state calculations by fast computing machines},}\ }\href
  {\doibase 10.1063/1.1699114} {\bibfield  {journal} {\bibinfo  {journal} {The
  Journal of Chemical Physics}\ }\textbf {\bibinfo {volume} {21}},\ \bibinfo
  {pages} {1087--1092} (\bibinfo {year} {1953})}\BibitemShut {NoStop}%
\bibitem [{\citenamefont {Hastings}(1970)}]{HASTINGS01041970}%
  \BibitemOpen
  \bibfield  {author} {\bibinfo {author} {\bibfnamefont {W.~K.}\ \bibnamefont
  {Hastings}},\ }\bibfield  {title} {\enquote {\bibinfo {title} {{Monte Carlo
  sampling methods using Markov chains and their applications}},}\ }\href
  {\doibase 10.1093/biomet/57.1.97} {\bibfield  {journal} {\bibinfo  {journal}
  {Biometrika}\ }\textbf {\bibinfo {volume} {57}},\ \bibinfo {pages} {97--109}
  (\bibinfo {year} {1970})}\BibitemShut {NoStop}%
\bibitem [{\citenamefont {Crowley}\ and\ \citenamefont
  {Green}(2016)}]{PhysRevA.94.062106}%
  \BibitemOpen
  \bibfield  {author} {\bibinfo {author} {\bibfnamefont {Philip J.~D.}\
  \bibnamefont {Crowley}}\ and\ \bibinfo {author} {\bibfnamefont {A.~G.}\
  \bibnamefont {Green}},\ }\bibfield  {title} {\enquote {\bibinfo {title}
  {Anisotropic landau-lifshitz-gilbert models of dissipation in qubits},}\
  }\href {\doibase 10.1103/PhysRevA.94.062106} {\bibfield  {journal} {\bibinfo
  {journal} {Phys. Rev. A}\ }\textbf {\bibinfo {volume} {94}},\ \bibinfo
  {pages} {062106} (\bibinfo {year} {2016})}\BibitemShut {NoStop}%
\bibitem [{Note2()}]{Note2}%
  \BibitemOpen
  \bibinfo {note} {An analogous noise-induced bias was observed for the problem
  in Ref.~\cite {q-sig2}}\BibitemShut {NoStop}%
\bibitem [{\citenamefont {Vidal}\ and\ \citenamefont
  {Werner}(2002)}]{Vidal:02a}%
  \BibitemOpen
  \bibfield  {author} {\bibinfo {author} {\bibfnamefont {G.}~\bibnamefont
  {Vidal}}\ and\ \bibinfo {author} {\bibfnamefont {R.~F.}\ \bibnamefont
  {Werner}},\ }\bibfield  {title} {\enquote {\bibinfo {title} {Computable
  measure of entanglement},}\ }\href
  {http://link.aps.org/doi/10.1103/PhysRevA.65.032314} {\bibfield  {journal}
  {\bibinfo  {journal} {Phys. Rev. A}\ }\textbf {\bibinfo {volume} {65}},\
  \bibinfo {pages} {032314--} (\bibinfo {year} {2002})}\BibitemShut {NoStop}%
\bibitem [{\citenamefont {Biamonte}\ and\ \citenamefont
  {Love}(2008)}]{Biamonte:07}%
  \BibitemOpen
  \bibfield  {author} {\bibinfo {author} {\bibfnamefont {Jacob~D.}\
  \bibnamefont {Biamonte}}\ and\ \bibinfo {author} {\bibfnamefont {Peter~J.}\
  \bibnamefont {Love}},\ }\bibfield  {title} {\enquote {\bibinfo {title}
  {Realizable hamiltonians for universal adiabatic quantum computers},}\ }\href
  {\doibase 10.1103/PhysRevA.78.012352} {\bibfield  {journal} {\bibinfo
  {journal} {Phys. Rev. A}\ }\textbf {\bibinfo {volume} {78}},\ \bibinfo
  {pages} {012352} (\bibinfo {year} {2008})}\BibitemShut {NoStop}%
\bibitem [{\citenamefont {{E.B. Davies}}(1976)}]{Davies:76}%
  \BibitemOpen
  \bibfield  {author} {\bibinfo {author} {\bibnamefont {{E.B. Davies}}},\
  }\href@noop {} {\emph {\bibinfo {title} {{Quantum Theory of Open Systems}}}}\
  (\bibinfo  {publisher} {{Academic Press}},\ \bibinfo {address} {London},\
  \bibinfo {year} {1976})\BibitemShut {NoStop}%
\bibitem [{\citenamefont {Lindblad}(1976)}]{Lindblad:76}%
  \BibitemOpen
  \bibfield  {author} {\bibinfo {author} {\bibfnamefont {G.}~\bibnamefont
  {Lindblad}},\ }\bibfield  {title} {\enquote {\bibinfo {title} {On the
  generators of quantum dynamical semigroups},}\ }\href {\doibase
  10.1007/BF01608499} {\bibfield  {journal} {\bibinfo  {journal} {Comm. Math.
  Phys.}\ }\textbf {\bibinfo {volume} {48}},\ \bibinfo {pages} {119--130}
  (\bibinfo {year} {1976})}\BibitemShut {NoStop}%
\end{thebibliography}
%

\appendix 
\section{Master equation} \label{App:ME}
%
We provide details of the weak-coupling limit master equation. While the equation can be derived under suitable assumptions from a microscopic model \cite{ABLZ:12-SI}, here we assume the master equation is a complete open-system description of our system. The master equation is a time-dependent Davies master equation \cite{Davies:76} that is in Lindblad form \cite{Lindblad:76}, and we assume identical and independent baths on each qubit:
\begin{eqnarray}
\frac{d}{dt} \rho(t) &=& -i \left[ \frac{H(t)}{\hbar}, \rho(t) \right] + \sum_{i=1}^n \sum_{w} \gamma(w) \times \nonumber \\ 
&& \hspace{-1.5cm}  \left[ L_{w,i}(t) \rho(t) L_{w,i}(t)^\dagger  - \frac{1}{2} \left\{ L_{w,i}^{\dagger}(t) L_{w,i}(t), \rho(t) \right\} \right] \ ,
\end{eqnarray}
where the index $i$ runs over the qubits and the index $\omega$ runs over all possible energy eigenvalue differences of the Hamiltonian (the Bohr frequencies). The Lindblad operators are given by
\beq
L_{w,i}(t) = \sum_{a,b} \delta_{w, E_b(t) - E_a(t)} \bra{E_a(t)} \sigma^z_i \ket{E_b(t)} \ket{E_a(t)}\bra{E_b} \ ,
\eeq
where we have taken a dephasing system-bath interaction on each qubit. We assume an Ohmic bath, such that we have:
\beq
\gamma(w) = \frac{2 \pi \kappa^2 w e^{-|w| / w_c}}{1 - e^{-\frac{\beta}{\hbar \omega} w}} \ ,
\eeq
where we take $w_c = 8 \pi$, and $\kappa^2$ is the dimensionless system-bath coupling. 

\begin{figure}[tbhp]
\begin{center}
\subfigure[]{\includegraphics[width=0.48\columnwidth]{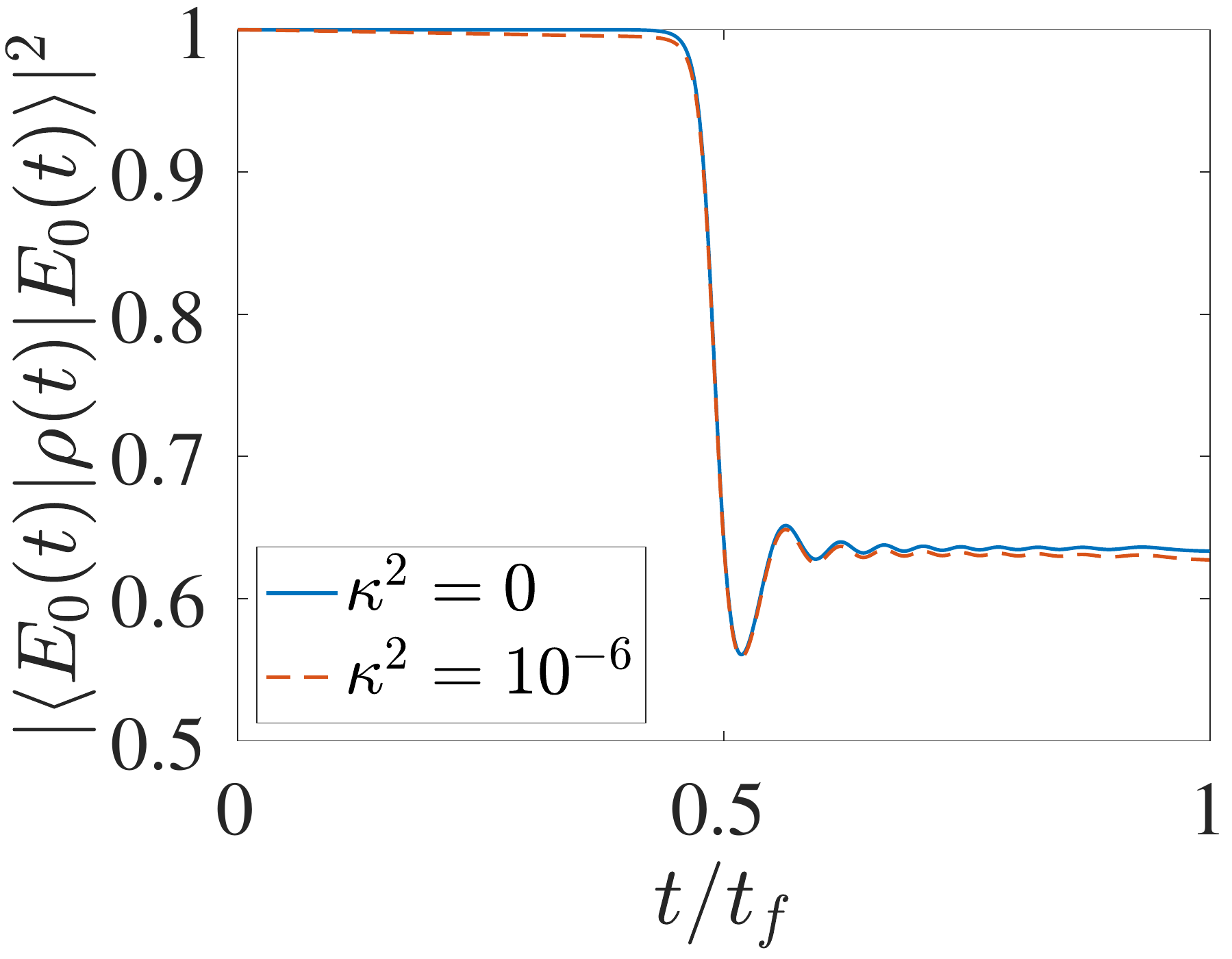} \label{fig:InstantGSa}}
\subfigure[]{\includegraphics[width=0.48\columnwidth]{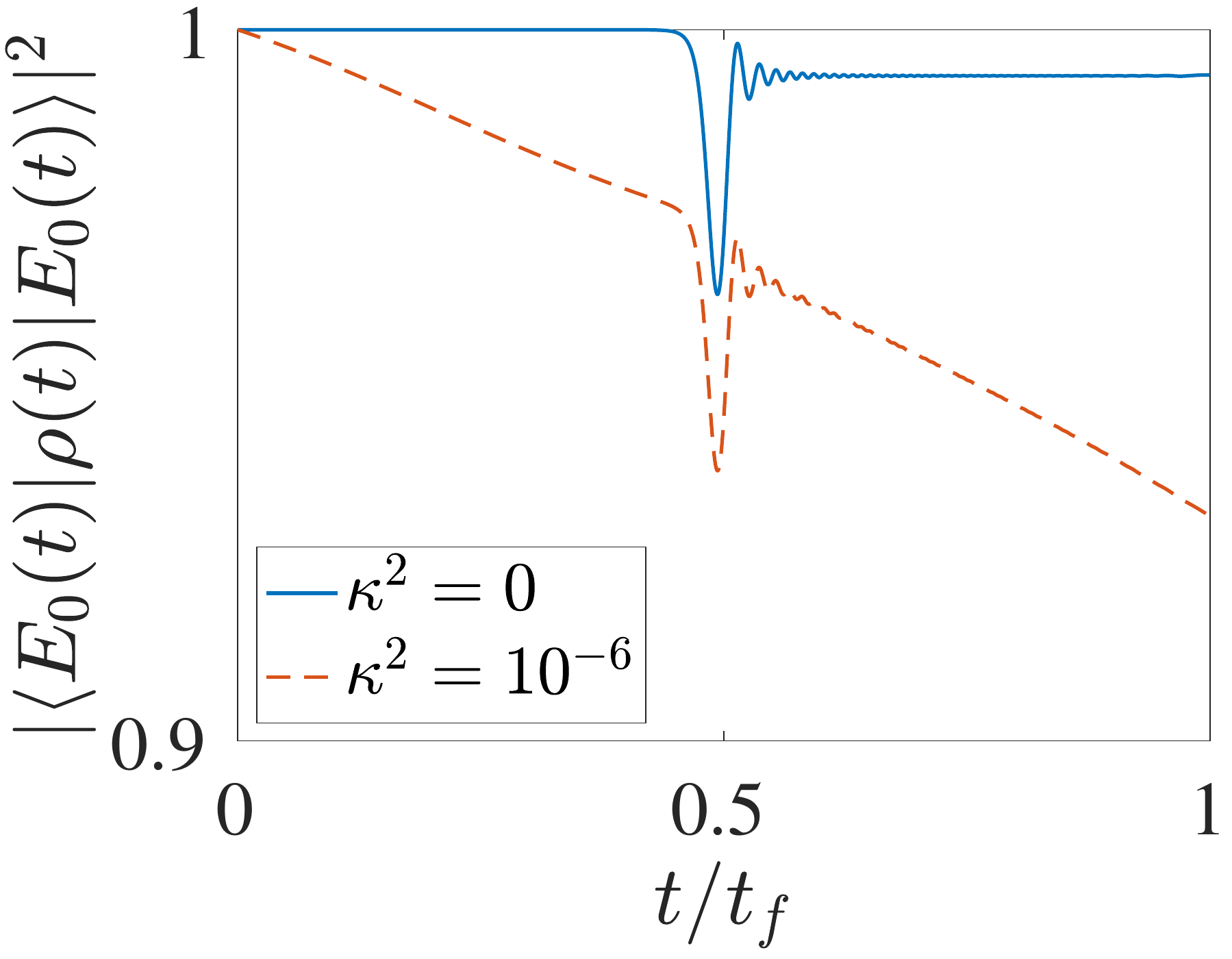} \label{fig:InstantGSb}}
   \caption{Instantaneous ground state population along the anneal according to the weak-coupling master equation simulations with $\alpha = 2, \beta = 0.05$, dimensionless system-bath coupling $\kappa^2$, energy scale $k_B T / \hbar \omega = 1.57$ and (a) $\omega t_f = 10^3$ and (b) $\omega t_f = 5 \times 10^3$.}  \label{fig:InstantGS}
\end{center}
\end{figure}

We give examples of the evolution in Fig.~\ref{fig:InstantGS} for two different annealing times.  We see that depending on the value of $\kappa^2$ relative to the annealing time, the relevant error can be either the unitary non-adiabatic transitions (Fig.~\ref{fig:InstantGSa}) or thermal transitions out of the ground state (Fig.~\ref{fig:InstantGSb}).  Note that in the latter case, we observe thermal excitations almost immediately after the start of the anneal.  This is because we have picked a rather high temperature energy scale relative to the overall energy of our Hamiltonian.
%
\section{Spin-Vector Monte Carlo} \label{App:SVMC}
The algorithm proceeds as follows.  In order to ensure that points on the sphere are picked uniformly, we pick new spin-vector with angles given by:
\beq
\theta = \cos^{-1} (2 v -1) \ , \quad \phi = 2 \pi u \ ,
\eeq
where $(u,v)$ are two uniform random variables on [0,1). We calculate the energy difference $\Delta E$ according to the potential energy in Eq.~\eqref{eqt:SVMCPotential} in the main text.  If $\Delta E < 0$, the new spin-vector is accepted; if $\Delta E >0$, then we draw another uniform random number on (0,1) $\eta$, and if $\eta < \exp(-\beta \Delta E)$, then the new spin vector is accepted.  Otherwise, no update is performed.  We discretize our $s$ parameter into $n_{\mathrm{sw}}$ identical steps from 0 to 1, and for each step we perform a single such update attempt on each spin.  Therefore, our Monte Carlo algorithm performs $n_{\mathrm{sw}}$ sweeps, performing a single sweep for each increment of the discretized $s$ value, corresponding to a total of $2 n_{\mathrm{sw}}$ single-spin update attempts.
\end{document}